\documentclass[aps,prb,twocolumn,superscriptaddress,eqsecnum,showpacs,showkeys,amsmath,amssymb]{revtex4}

\usepackage{amsfonts}
\usepackage{amssymb,amsmath}
\usepackage{graphicx}
\usepackage{dcolumn}
\usepackage{bm}
\usepackage{color}
\DeclareMathAlphabet\mathbfcal{OMS}{cmsy}{b}{n}

\begin{document}

\title{Spin-1/2 $XY$ chain magnetoelectric: effect of zigzag geometry}

\author{Ostap Baran}
\affiliation{Institute for Condensed Matter Physics,
          National Academy of Sciences of Ukraine,
          Svientsitskii Street 1, 79011 L'viv, Ukraine}

\author{Vadim Ohanyan}
\affiliation{Department of Theoretical Physics,
          Yerevan State University,
          Alex Manoogian 1, 0025 Yerevan, Armenia}
\affiliation{Joint Laboratory of Theoretical Physics
             ICTP Affiliated centre in Armenia,  2. Alikhanian Br. Street, Yerevan, Armenia, 0036
}

\author{Taras Verkholyak}
\affiliation{Institute for Condensed Matter Physics,
          National Academy of Sciences of Ukraine,
          Svientsitskii Street 1, 79011 L'viv, Ukraine}



\date{\today}

\pacs{75.10.Jm, 
      75.10.-b, 
      05.50.+q  
      }

\keywords{magnetoelectric effect,
          Katsura-Nagaosa-Balatsky mechanism,
          spin-1/2 $XY$ chains}

\begin{abstract}
A spin-1/2 $XY$ chain model of magnetoelectric on  a zigzag chain is considered rigorously. The magnetoelectric coupling is described within the Katsura-Nagaosa-Balatsky mechanism. In the zigzag geometry it leads to the staggered Dzyaloshinskii-Moriya interaction.
By non-uniform spin-rotations the model is reduced to a dimerized $XY$ chain and solved exactly using the Jordan-Wigner transformation. We analyze the ground-state phase diagram of the model, zero and finite temperature magnetoelectric effect, obtain the magnetization and polarization curves versus magnetic and electric fields, as well as the parameters of anisotropic dielectric and magnetoelectric response.
It is also shown that the electric field may enhance the magnetocaloric effect in the model.
\end{abstract}

\maketitle

\section{Introduction}
\label{sec1}
Among multiferroics, the materials simultaneously exhibiting more than one ferroic order
\cite{dong15,mee1,mee2,app1,app2}, magnetoelectrics play a special role, due to their broad and important technical applications\cite{dong15}. Magnetoelectric effect (MEE) is, in general, the term for denoting the vast class of phenomena of intercoupling of magnetization and polarization in matter\cite{dong15,mee1,mee2}. The most common manifestation of the MEE in solids  is the magnetization dependence on the electric field and polarization dependence on the magnetic field. The magnetoelectric materials in general and the spin related ferroelectricity are particularly important for possible application in various electronic and spintronic devices\cite{app1,app2,wan14, ort15,mat15}. Moreover, there are most recent results evidencing the possibility to generate a field of a magnetic monopole by placing an electric charge on the surface of a linear magnetoelectric slab\cite{mei18}. There exist several physical mechanisms coupling the local magnetic moments of the magnetic material with the local polarization of the unit cell. The one to be considered in the present paper is based on the so-called spin current model or inverse Dzyaloshinskii-Moriya (DM) model and is referred to as the Katsura-Nagaosa-Balatsky (KNB) mechanism\cite{KNB1,KNB2}. The KNB mechanism\cite{KNB1,KNB2} links the dielectric polarization corresponding to the pair of spins at the adjacent lattice sites, with the spin current across the bond given by the following expression:
\begin{eqnarray}
\label{201}
\mathbf{P}_{ij}= \gamma\mathbf{e}_{ij}\times\mathbf{s}_i\times\mathbf{s}_j,
\end{eqnarray}
where $\mathbf{e}_{ij}$ is the unit vector pointing from site $i$ to site $j$, and
$\gamma$ is the coefficient that connects the electric polarization with the magnetic current operator.

Several exact results are known on the magnetoelectric models with KNB mechanism\cite{KNB1,KNB2}: the spin-1/2 $XXZ$ chain\cite{vadim}, the spin-1/2 $XY$ chain with three-spin interaction\cite{mench15, szn18}, generalized quantum compass model with magnetoelectric coupling \cite{oles}.
The results of Refs. [\onlinecite{vadim, mench15}] were further confirmed in Refs. [\onlinecite{thakur, thakur2}]. The link between DM-terms and the quantum phase transitions of a generalized compass chain with staggered Dzyaloshinskii-Moriya interaction
was also considered recently \cite{wu17}.

 There are a  number of real magnetic materials with one-dimensional or quasi-one dimensional magnetic lattice in which the MEE is realized according to the KNB mechanism\cite{LiCu2O21,LiCu2O22,LiCu2O23,LiCu2O24,LiCuVO41,LiCuVO42,LiCuVO43,CuCl2,CuBr2,CuX2}. For materials like LiCu$_{2}$O$_2$\cite{LiCu2O21,LiCu2O22,LiCu2O23,LiCu2O24}, LiCuVO$_4$\cite{LiCuVO41,LiCuVO42,LiCuVO43}, copper halides\cite{CuCl2,CuBr2,CuX2} and others it is believed that more or less adequate model describing the MEE is believed to be the so-called multiferroic spin chain (MSC), $S=1/2$ quantum spin chain with competing ferromagnetic nearest-neighbor and antiferromagnetic next-nearest-neighbor interactions:
 \begin{eqnarray}
 \label{J1J2}
 \mathcal{H}&=&J_1\sum_{j=1}^N\mathbf{s}_{j}{\cdot}\mathbf{s}_{j+1}+J_2\sum_{j=1}^N\mathbf{s}_{j}{\cdot}\mathbf{s}_{j+2}\\
   &-&{\mathbf E}{\cdot}\mathbf{P}-\mathbf{B}{\cdot}\mathbf{M}. \nonumber
 \end{eqnarray}
Here electric (magnetic) field vector, ${\mathbf E}$ ($\mathbf{B}$), is coupled to polarization (magnetization) given by
\begin{eqnarray}
&&\mathbf{P}=\gamma\sum_{j=1}^N\left(\mathbf{s}_j\;s_{j+1}^x-s_j^x\;\mathbf{s}_{j+1}\right),\\
&&\mathbf{M}=g\mu_B\sum_{j=1}^N\mathbf{s}_j, \nonumber
\end{eqnarray}
respectively. Here,
$g$ is the g-factor of a magnetic ion, and $\mu_B$ is the Bohr magneton.
The chain is supposed to have strictly linear form in the $x$ direction. Besides the MEE by itself the MSC recently received a considerable amount of attention from various other contexts, e.g. quantum information processing\cite{azi14a}, quantum Otto cycles\cite{azi14b}, pulse and quench dynamics\cite{azi16}, many-body localization\cite{sta17} etc.  It worth mentioning, that the physics of the MSC is very rich and complicated even without the electric field\cite{hei06,vek07,fur10,sat11a,sat11b,kol12}. However, by virtue of its complexity the model (\ref{J1J2}) allows only numerical treatment. Nevertheless, the exact solutions of the simplified spin models demonstrating the MEE due to KNB mechanism are very important as they can shed light on the general universal properties of the magnetoelectrics and offer a unique opportunity to figure out their general features analytically\cite{vadim, mench15, szn18, oles, thakur, thakur2, wu17}. In one of the previous works the integrable model of the $S=1/2$ $XXZ$ chain with DM-interaction has been considered as a model of the linear spin-chain with the KNB mechanism\cite{vadim}. MEE in this system has shown to be trivial, which means the absence of polarization (magnetization) at zero external electric (magnetic) field. However, the magnetic (electric) field affects the polarization (magnetization) when the electric (magnetic) field is on. The next exactly solvable linear spin chain model with KNB mechanism, $XX$ chain with three-spin interactions\cite{mench15,szn18} demonstrates non trivial MEE, i. e. only magnetic (electric) field can induce polarization (magnetization). This takes place due to three-spin terms, which mimics a microscopic interaction between local magnetic moment and local polarization.

 In the present paper we continue our research on exactly solvable spin models with KNB mechanism. However, as the form of the local polarization is essentially dependent on the geometry of the exchange interaction bonds between the spins, here we consider the effects of non-uniform local polarization throughout the chain. In the simplest case the local polarization for the bonds has period two. Within the KNB mechanism this can be the case if one consider the chain to be folded to form a zigzag. Thus, formally, we deal with the $XX$ model with alternating DM-terms in magnetic field. With the aid of the Jordan-Wigner transformation the system is mapped into the free spinless fermions. We studied in  detailed the zero- and finite-temperature magneto-thermal and magnetoelectric properties of the model.

Although, the model we study in the present work finds no exact realization among the multiferroic materials known at the moment, the exact treatment and analytical results obtained in the paper can shed light on the general and universal features of the MEE in case of the staggered local polarization as well as on the influence of the zigzag geometry of the bonds to the KNB-mechanism. For instance, we obtained several universal results about the direction of the total polarization vector which can be easily generalized to the more realistic quantum spin chain models (see Appendix~\ref{app:equal_angles} and Appendix~\ref{app:angles_small_fields}).  Moreover, due to the ongoing progress in the material science, the relevance of the model considered in the present paper will be possible to check in novel materials in the near future.

  The paper is organized as follows: in the second Section we describe the KNB mechanism for the zigzag geometry, the next, third, section presents the exact solution for the $XY$ model of a magnetoelectric on a zigzag chain, in the fourth Section we describe its ground state properties, including the zero-temperature MEE, the next, fifth Section devoted to the finite-temperature properties of the model and MEE, the last sixth Section concludes the paper.

\section{KNB mechanism for spin-chains: the influence of the geometry}
\label{sec2}
 The form of the expression for the dielectric polarization in terms of the spin operators for the magnetoelectric material with KNB mechanism essentially depends on the geometry of the magnetic unit cell.
 For the linear chain in direction $x$ ($\mathbf{e}_{ij}\equiv\mathbf{e}_x$) the corresponding expressions for the polarization vector components can be recovered from Eq.~(\ref{201}) and are quite simple\cite{vadim,mench15,szn18}:
\begin{eqnarray}
\label{202}
&&P_{j,j+1}^x=0,\\
&&P_{j,j+1}^y=\gamma (s_j^ys_{j+1}^x-s_{j}^xs_{j+1}^y),\nonumber\\
&&P_{j,j+1}^z=\gamma (s_j^zs_{j+1}^x-s_{j}^xs_{j+1}^z).\nonumber
\end{eqnarray}
However, even small changes in the spatial arrangement of the spins can bring sufficient complication of the structure of local polarization. For instance, one can consider the spin chain laying in the $xy$ plain but with possibility of the arbitrary planar angle $\theta_{j}$ between the $x$-axis and $j$-th bond connecting $j$-th and $(j+1)$-th site.
Then, the expression for the dielectric polarization corresponding to $j$-th bond should be modified according to the KNB formula, Eq.~(\ref{201}), and an altered direction of the bond given by unit vector
$\mathbf{e}_{j,j+1}=\cos\theta_j\;\mathbf{e}_x+\sin\theta_j\;\mathbf{e}_y$:
\begin{eqnarray}\label{Pxy}
\mathbf{P}_{j,j+1}=\gamma\left(\cos\theta_j\;\mathbf{e}_x+\sin\theta_j\;\mathbf{e}_y\right)\times\mathbf{s}_j\times\mathbf{s}_{j+1}.
\end{eqnarray}
The corresponding components of the polarization are
 \begin{eqnarray}\label{Psc}
P_{j,j+1}^x&=&\gamma\sin\theta_j\left(s_j^xs_{j+1}^y-s_j^ys_{j+1}^x\right),\\
P_{j,j+1}^y&=&-\gamma\cos\theta_j\left(s_j^xs_{j+1}^y-s_j^ys_{j+1}^x\right),\nonumber\\
P_{j,j+1}^z&=&\gamma\cos\theta_j\left(s_j^zs_{j+1}^x-s_j^xs_{j+1}^z\right)\nonumber\\
&+&\sin\theta_j\left(s_j^zs_{j+1}^y-s_j^ys_{j+1}^z\right).\nonumber
 \end{eqnarray}
 Considering the constant and homogeneous electric field laying in the $xy$ plain,
 ${\mathbf{E}}=\left({ E}_x, { E}_y,0\right)=\left({ E}\cos\varphi^{}_{\rm E}, { E}\sin\varphi^{}_{\rm E},0\right)$,
 where $\varphi^{}_{\rm E}$ is the angle between the
 $x$-axis and electric field vector, $E=\sqrt{{ E}_x^2+{ E}_y^2}$,
 one can write down the energy contribution of the interaction between the dipole
 moment of the chain and electric field in the following form:
 \begin{eqnarray}\label{EPc}
 -{\mathbf{E}}{\cdot}\mathbf{P}=-E\gamma\sum_{j=1}^N\sin\left(\theta_j-\varphi^{}_{\rm E}\right)\left(s_j^xs_{j+1}^y-s_j^ys_{j+1}^x\right).
 \end{eqnarray}

\begin{figure}[tbp]
 \begin{center}
  \includegraphics[width=0.73\columnwidth]{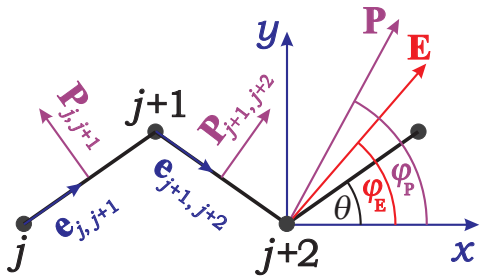}
 \end{center}
\caption{The zigzag chain with indicated system axes, electric field vector
$\mathbf{E}$, polarization $\mathbf{P}$, bond polarization $\mathbf{P}_{j,j+1}$
and unit vector $\mathbf{e}_{j,j+1}$ pointing from $j$th site to $(j+1)$th site.
Here the $z$-component of the bond polarization is equal zero, because
$\mathbf{E}=(E_x,E_y,0)$. }
\label{fig_zigzag}
\end{figure}

 Let us now consider the same spin chain but bent to form a zigzag with the fixed equal angles between each bonds (see Fig.~\ref{fig_zigzag}).
 Therefore, the angle $\theta_j=(-1)^{j+1}\theta$ is staggering with respect to $x$-axis and
 the polarization vectors between odd-even (bottom-top) and even-odd (top-bottom) pairs of spin now are different and are given by the following expressions:
\begin{eqnarray}
\label{203}
\mathbf{P}_{j,j+1}=\gamma (\cos\theta\mathbf{e}_x-(-1)^j\sin\theta\mathbf{e}_y)\times\mathbf{s}_{j}\times\mathbf{s}_{j+1},
\end{eqnarray}
Thus, Eq.~(\ref{EPc}) can be explicitly written as follows
\begin{eqnarray}
\label{205}
{-}{\mathbf{E}}{\cdot}\mathbf{P}{=}
\gamma\sum_{j=1}^N\left({ E}_y\cos\theta{+}({-}1)^j { E}_x\sin\theta\right)\left(s_{j}^xs_{j+1}^y {-} s_{j}^ys_{j+1}^x\right).\nonumber\\
\end{eqnarray}

\section{Model and exact solution}
\label{sec3}
We consider the quantum spin-1/2 $XY$ model of $N$ spins in the electric and magnetic field on the zigzag chain (see Fig.~\ref{fig_zigzag}) described by the following Hamiltonian:
 \begin{widetext}
\begin{eqnarray}
\label{301}
\mathcal{H}&=&\sum_{j=1}^N\left(J(s_j^x s_{j+1}^x+s_j^y s_{j+1}^y)+\gamma \left(E_y\cos\theta+(-1)^jE_x\sin\theta \right)\left(s_j^x s_{j+1}^y-s_j^y s_{j+1}^x\right)\right)-h\sum_{j=1}^Ns_j^z
\end{eqnarray}
\end{widetext}
Here, we introduced the renormalized magnetic field $h=g\mu_B B_z$.
First term in the Hamiltonian (\ref{301}) refers to the superexchange coupling between neighboring spins, second [third] terms corresponds to the energy of the model in the electric ${\mathbf E}=(E_x, E_y, 0)$ [magnetic $(0,0,h)$] field.
It can be noticed that the effect of the electric field discussed in the previous section is given by the staggered DM interaction terms.
From Eqs.~(\ref{205}), (\ref{301}), we see that the coefficient $\gamma$ appears in the term of the electric field only. Here, we set $\gamma=1$ bearing in mind
that one has to use its specific value corresponding to each  particular case of real materials. Thus, by comparing to the experiment our results for the polarization should be multiplied by $\gamma$, while the electric field needs to be divided by $\gamma$.
In all further calculations we also set $J=1$ and
restrict ourselves to $0<\theta<\pi/3$, since for larger values the distance between next-nearest neighbors become shorter than for the nearest neighbors.
It should be mentioned that the model (\ref{301}) resembles the two-sublattice $XY$ chain with the DM interaction studied in Ref. [\onlinecite{zvyagin92}], while the quantum compass model with the staggered DM interaction has been considered recently  in Ref. [\onlinecite{wu17}].

Here we face the case of isotropic $XY$ interaction, where
the Hamiltonian can be further simplified by the rotation transformation in the $xy$
plain \cite{oshikawa97,bocquet01,perk76,derzhko00,derzhko06}:
\begin{eqnarray}
\label{rotation_transf}
\tilde{s}_{j}^x&=&{s}_{j}^x\cos\phi_j + {s}_{j}^y\sin\phi_j,
\nonumber\\
\tilde{s}_{j}^y&=&-{s}_{j}^x\sin\phi_j + {s}_{j}^y\cos\phi_j,
\nonumber\\
\tilde{s}_{j}^z&=&{s}_{j}^z,
\end{eqnarray}
where $\phi_{2j}=(j-1)(\phi^+ + \phi^-)+\phi^-$, $\phi_{2j+1}=j(\phi^+ + \phi^-)$, and
$\tan\phi^{\pm}=E_{\pm}$. Hereinafter we use the notations
$E_{\pm}=E_y\cos\theta\pm E_x\sin\theta \equiv E\sin(\varphi^{}_{\rm E}\pm\theta)$
and $E=\sqrt{E_x^2+E_y^2}$.
As a result we come to the dimerized $XY$ chain considered in Ref.~[\onlinecite{taylor85}]:
\begin{eqnarray}
\label{dim_ham}
\mathcal{H}&=&\sum_{j}\left[(J_+ + (-1)^j J_-)\left(\tilde{s}_j^x \tilde{s}_{j+1}^x+\tilde{s}_j^y \tilde{s}_{j+1}^y\right)
- h\tilde{s}_j^z \right],
\nonumber\\
J_{\pm}
&=&\frac{1}{2}\left(\sqrt{1+E_+^2}\pm \sqrt{1+E_-^2}\right).
\end{eqnarray}
It should be stressed that the described elemination of DM terms is possible only in the case of the nearest-neighbor interaction (see the discussions in Ref.~[\onlinecite{bocquet01}]).

Using the Jordan-Wigner transformation\cite{lsm} the model can be reduced to the noninteracting spinless fermion gas, and then brought to the diagonal form by Fourier and unitary transformation (see details in Ref.~[\onlinecite{taylor85}]):
\begin{eqnarray}
\label{fermi_ham}
\mathcal{H}&=&\sum_{-\pi<k\leq\pi}\Lambda_k\left(a_k^+ a_k -\frac{1}{2}\right),\\
\Lambda_k&=&{-}h{+}{\rm sign}(\cos k)\sqrt{J_+^2\cos^2k {+} J_-^2\sin^2k},\nonumber
\end{eqnarray}
where $a_k$ and $a_k^+$ are the Fermi annihilation and creation operators with quasimomentum $k=2\pi l/N$ ($l=-N/2+1,\dots,N/2$),
$\Lambda_k$ is the spectrum of the effective spinless fermion excitations of the dimerized model (\ref{dim_ham}).
The correspondence between the original spin model and its fermionic counterpart for linear $XY$ magnetoelectric can be found in Refs.~[\onlinecite{vadim,mench15}]. Fermions create magnon excitations in the Hamiltonian (\ref{301}) and the complete empty (filled) state corresponds to the fully polarized down (up) spin model.

We will take the thermodynamic limit in all further calculations.
The free energy per site can be easily found as
\begin{eqnarray}
\label{free_en}
f=-\frac{1}{\pi\beta}\int\limits_0^\pi dk \ln \left[ 2\cosh\left(\frac{\beta\Lambda_k}{2} \right) \right],
\end{eqnarray}
where $\beta=\frac{1}{k_{\rm B} T}$ is the inverse temperature $T$, and $k_{\rm B}$ is the Boltzmann constant.
All other thermodynamic quantities of the system can be also readily obtained by differentiating Eq.~(\ref{free_en}), i.e. the magnetization per site
\begin{eqnarray}
m_z &=& \frac{1}{N}\sum_{j=1}^N\langle s_i^z\rangle
=-\frac{1}{2\pi}\int\limits_0^\pi dk \tanh\left(\frac{\beta\Lambda_k}{2} \right),
\end{eqnarray}
the $x-$ and $y-$components of the electric polarization per site
\begin{eqnarray}
\label{el_pol}
p_\mu &=& \frac{1}{N}\sum_{j=1}^N\langle P_{j,j+1}^\mu\rangle
\nonumber\\
&&= \frac{1}{2\pi}\int\limits_0^\pi dk\left( J_+\frac{\partial J_+}{\partial E_\mu}\cos^2k + J_-\frac{\partial J_-}{\partial E_\mu}\sin^2k \right)
\nonumber\\
&&\times \frac{{\rm sign}(\cos k)\tanh\left(\frac{\beta\Lambda_k}{2} \right)}{\sqrt{J_+^2\cos^2k+J_-^2\sin^2k}}, \;\; \mu=x,y,
\\
\label{dJdE}
\frac{\partial J_{\pm}}{\partial E_x}&=&
\frac{\sin\theta}{2}\left(
\frac{E_+}{\sqrt{1+E_+^2}}
\mp
\frac{E_-}{\sqrt{1+E_-^2}}
\right),
\nonumber\\
\frac{\partial J_{\pm}}{\partial E_{y}}&=&
\frac{\cos\theta}{2}\left(
\frac{E_+}{\sqrt{1+E_+^2}}
\pm
\frac{E_-}{\sqrt{1+E_-^2}}
\right),
\end{eqnarray}
the entropy per site
\begin{eqnarray}
S &=&\frac{k_{\rm B}}{\pi}
{\int\limits_0^\pi} dk\left\{ \ln \left[ 2\cosh\left( \!\frac{\beta\Lambda_k}{2} \!\right) \right]
{-}\frac{\beta \Lambda_k}{2}\tanh\left( \! \frac{\beta\Lambda_k}{2} \! \right)\right\},
\nonumber\\
\end{eqnarray}
and the specific heat per site
\begin{eqnarray}
c=\frac{k_{\rm B}\beta^2}{4\pi}\int\limits_0^\pi dk \frac{\Lambda_k^2}{\cosh^2\left(\frac{\beta\Lambda_k}{2} \right)}.
\end{eqnarray}
The important physical quantity to characterize the MEE is the magnetoelectric susceptibility (magnetoelectric tensor), given by the following expression:
\begin{eqnarray}
\alpha_{\mu\nu}=\left(\frac{\partial m_{\mu}}{\partial E_{\nu}}\right)_{T,\mathbf{B}}=\left(\frac{\partial P_{\nu}}{\partial h_{\mu}}\right)_{T,\mathbf{E}}.
\end{eqnarray}
In our case we have two components $\alpha_{z\mu}$ ($\mu=x,y$) of the magnetoelectric tensor:
\begin{widetext}
\begin{eqnarray}
\label{me_tensor}
&&\alpha_{z\mu}=-\frac{\beta}{4\pi}\int\limits_0^\pi dk\frac{{\rm sign}(\cos k)\left[J_+\frac{\partial J_+}{\partial E_\mu}\cos^2 k+J_-\frac{\partial J_-}{\partial E_\mu}\sin^2 k\right]}{\cosh^2\left(\frac{\beta\Lambda_k}{2}\right)\sqrt{J_+^2\cos^2 k+J_-^2\sin^2 k}}.
\end{eqnarray}
\end{widetext}

Looking at the transformed Hamiltonian (\ref{dim_ham}) one can notice some symmetries in the model. If the electric field is directed along only the $y$-axis, the system evidently does not feel any zigzag deformation and behaves identically to the quantum $XY$ model on a simple linear chain\cite{vadim, mench15}. Interestingly, the direction of field along $x$-axis, which corresponds to the staggered DM term, recovers the same limit of the linear chain. It is easy to check that in the former case $E_x\neq 0$, $E_y=0$ the Hamiltonian (\ref{301}) is explicitly dimerized, but the rotation (\ref{rotation_transf}) reduces it to the uniform form.
Additionally, in general case ($E_x\ne0$, $E_y\ne0$),
if we choose $\phi_{2j-1}=\phi_{2j}=2j\phi^{-}$  in the rotation transformation (\ref{rotation_transf}),
we get the following relation between the transformed and initial Hamiltonians
$\tilde{\mathcal H}(E_x\sin\theta,E_y\cos\theta)={\mathcal H}(E_y\cos\theta,E_x\sin\theta)$.
Thus, some thermodynamic functions (e.g., $f$, $m_z$ and $c$) are invariant
under simultaneous exchanges $\theta \rightarrow \pi/2-\theta$
and $E_x \leftrightarrow E_y$, while some electric characteristics
are transformed according to simple relations (e.g., $p_x \leftrightarrow p_y$).
It is also clear that the replacement
$E_\mu \rightarrow -E_\mu$ changes also only some electric characteristics
(e.g., $p_\mu \rightarrow -p_\mu$).
Therefore without loss of generality we hereinafter restrict our investigation
to the intervals $0\le\theta\le\pi/4$ and $0\le\varphi^{}_{\rm E}\le\pi/2$.

\section{Ground-state properties}
\begin{figure}[tbp]
 \begin{center}
  \includegraphics[width=0.8\columnwidth]{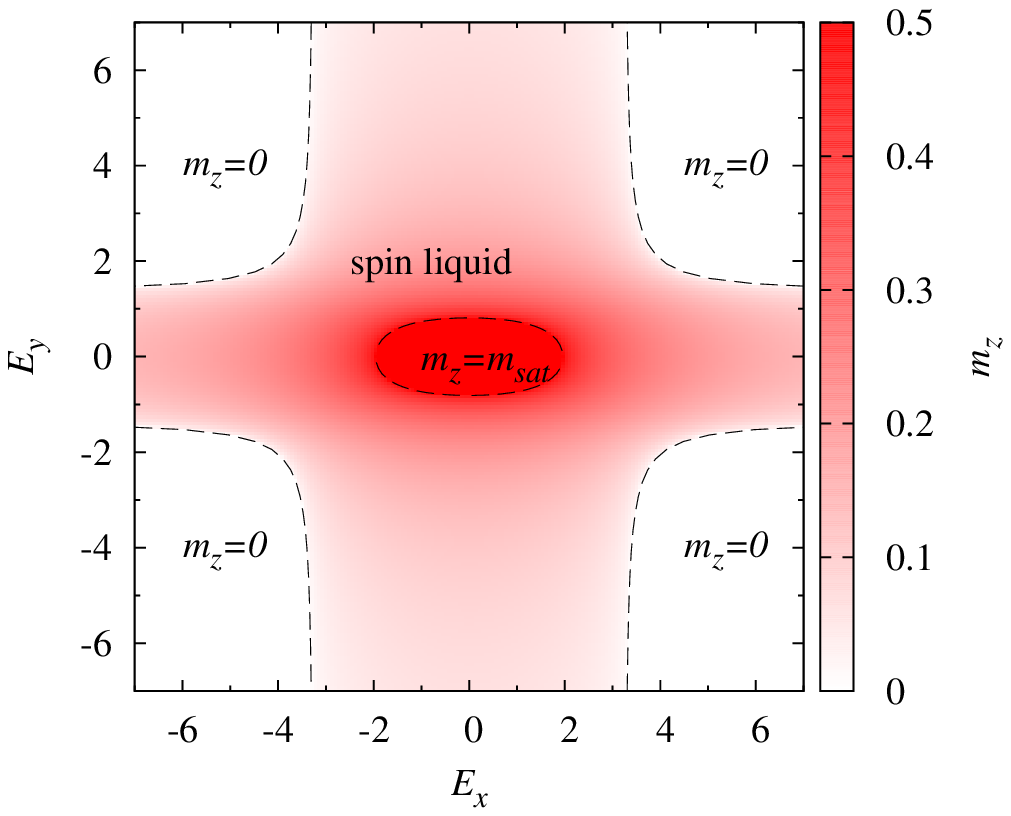}
  \includegraphics[width=0.8\columnwidth]{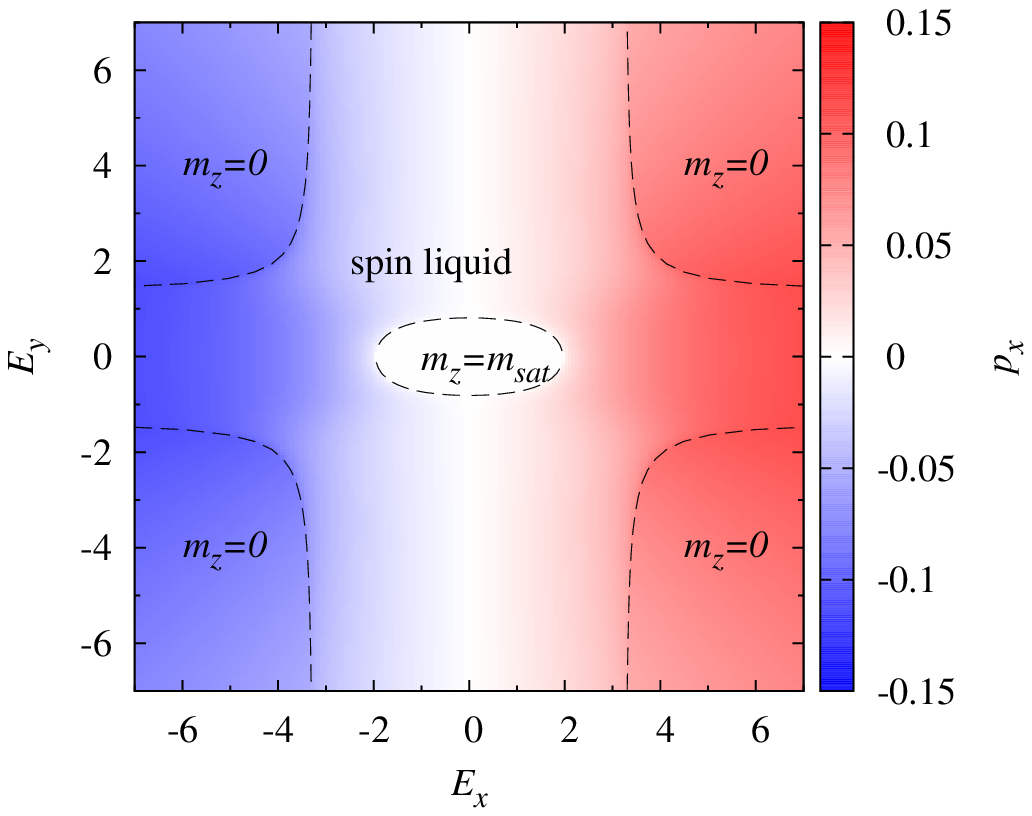}
  \includegraphics[width=0.8\columnwidth]{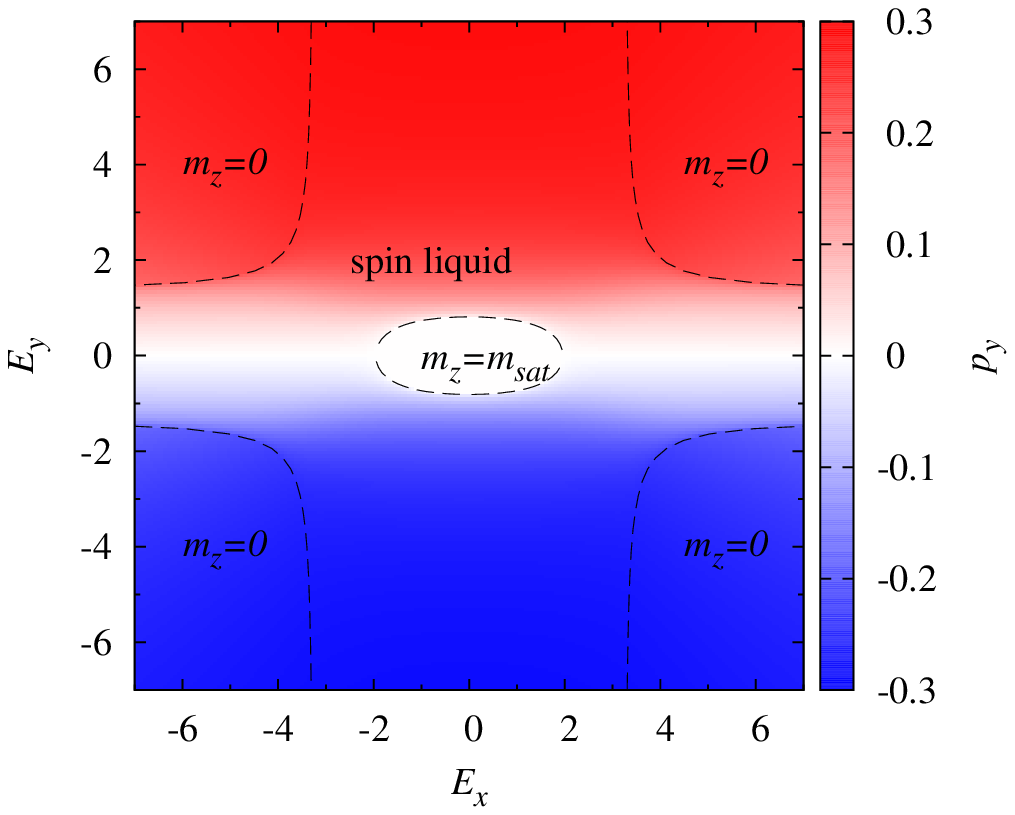}
 \end{center}
\caption{Density plots for the magnetization (upper panel) and polarizations $p_x$ and $p_y$ (middle and lower panels) at $T=0$ for $\theta=\pi/8$, $h=1.25$. Dashed lines indicate the boundaries between different ground-state phases.}
\label{fig_gs1}
\end{figure}

\begin{figure}[tbp]
 \begin{center}
  \includegraphics[width=0.8\columnwidth]{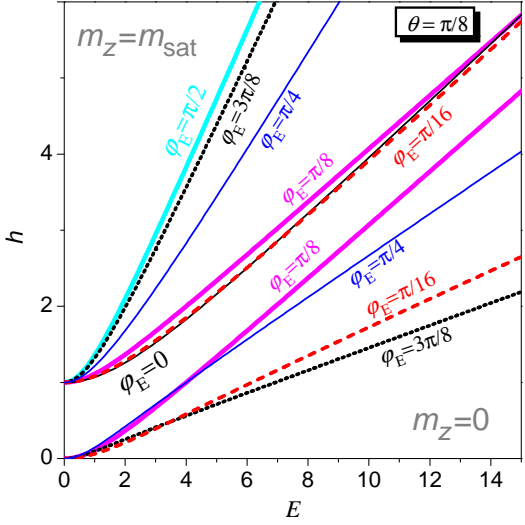}
 \end{center}
\caption{The ground state phase diagram for $\theta=\pi/8$ at different directions of the electric field  $\varphi^{}_{\rm E}=0,\pi/16,\pi/8,\pi/4,3\pi/8,\pi/2$.}
\label{fig_gs2}
\end{figure}

The application of the electric field in the $xy$ plane may transform the model into the effectively dimerized $XY$ chain (\ref{dim_ham}). As we show below it leads to the appearance of a new non-magnetic gapped phase and a new topology of the ground-state phase diagram.
For the sake of clarity, we choose $h\geq 0$.
From the analysis of the excitation spectrum $\Lambda_k$ (\ref{fermi_ham}), we deduce that the model is in the gapless spin-liquid phase if
$|J_-|< h < J_+$, since the excitation spectrum $\Lambda_k$ turns into zero at Fermi points
$\pm k_0$, given by the expression
\begin{eqnarray}
\label{f_point}
k_0=\arcsin\frac{\sqrt{1-(h/J_+)^2}}{\kappa},
\end{eqnarray}
where $\kappa=\sqrt{1-(J_-/J_+)^2}$. In case $h<|J_-|$ we have a gapped zero-plateau phase, while for $h>J_+$ all spins are directed along the magnetic field. The ground-state phase diagrams as functions of electric and magnetic fields are shown in Figs. \ref{fig_gs1}, and \ref{fig_gs2}.
One can see that the electric field destroys the saturated phase and drives the system to the
spin-liquid gapless phase. Finally, for rather strong fields in the $xy$ plain we obtained
a completely demagnetized phase $m_z=0$.
It should be noted that negative values
$E_x$ and $E_y$ are presented in Fig. \ref{fig_gs1}
for the sake of clarity.

The ground-state energy corresponding to the $T\to 0$ limit is expressed as:
\begin{eqnarray}
\label{gs_energy}
e_0=
\left\{
\begin{array}{lll}
-\frac{J_+}{\pi} {\mathbf E}(\kappa), & {\rm if}\; h\leq |J_-|,\\
-\left(\frac{1}{2}{-}\frac{k_0}{\pi}\right)h {-} \frac{J_+}{\pi} {\mathbf E}(k_0|\kappa), & {\rm if}\: |J_-|{\leq}h{\leq}J_+,\\
-\frac{h}{2}, & {\rm if}\; h\geq J_+,
\end{array}
\right.
\end{eqnarray}
where $k_0$ is the Fermi point given in Eq.~(\ref{f_point}),
${\mathbf E}(k_0|\kappa)=\int_0^{k_0}dk\sqrt{1-\kappa^2\sin^2 k}$ is the incomplete elliptic integral of the second kind of modulus $\kappa$, and ${\mathbf E}(\kappa)={\mathbf E}(\pi/2|\kappa)$.

The ground-state magnetization can be obtained to be
\begin{eqnarray}
\label{gs_mag}
m_z=
\left\{
\begin{array}{lll}
0, & {\rm if}\; h\leq |J_-|,\\
\frac{1}{2} - \frac{k_0}{\pi}, & {\rm if}\: |J_-|{\leq}h{\leq}J_+,\\
\frac{1}{2}, & {\rm if}\; h\geq J_+,
\end{array}
\right.
\end{eqnarray}

The expressions for the electric polarization $p_{\mu}=-\frac{\partial e_0}{\partial E_{\mu}}$ ($\mu=x,y$) can be derived by a straightforward calculation. In the saturated phase $h> J_+$ it equals zero. In the spin-liquid state ($|J_-|< h< J_+$) it is given by the following expression:
\begin{eqnarray}
\label{gs_pol1}
p_{\mu}&{=}&\frac{1}{\pi}\left\{
\frac{\partial J_+}{\partial E_{\mu}} {\mathbf E}(k_0|\kappa)
{+}\frac{J_+}{2\kappa^2}({\mathbf E}(k_0|\kappa) {-} {\mathbf F}(k_0|\kappa))\frac{\partial \kappa^2}{\partial E_{\mu}}
\right\},
\nonumber\\
\frac{\partial \kappa^2}{\partial E_{\mu}}&{=}&-\frac{2J_-}{J_+^2}
\left(\frac{\partial J_-}{\partial E_{\mu}}
-\frac{J_-}{J_+}\frac{\partial J_+}{\partial E_{\mu}}
\right),
\end{eqnarray}
where
${\mathbf F}(k_0|\kappa)=\int_0^{k_0}\frac{dk}{\sqrt{1-\kappa^2\sin^2 k}}$,
is the incomplete elliptic integrals of the first kind,
$\frac{\partial J_\pm}{\partial E_{\mu}}$ are given in Eq.~(\ref{dJdE}).
For the non-magnetic phase ($h<|J_-|$) we get
\begin{eqnarray}
\label{gs_pol2}
p_{\mu}
&{=}&\frac{1}{\pi}\left\{
\frac{\partial J_+}{\partial E_{\mu}} {\mathbf E}(\kappa)
{+} \frac{J_+}{2\kappa^2}({\mathbf E}(\kappa) {-} {\mathbf K}(\kappa))\frac{\partial \kappa^2}{\partial E_{\mu}}
\right\},
\end{eqnarray}
where
${\mathbf K}(\kappa)={\mathbf F}(\pi/2|\kappa)$.

The polarization is a non-analytic function in zero magnetic and electric fields. To show that, we can use an asymptotic expansion for the complete elliptic integral of the first kind for $\kappa\approx 1$:
\begin{eqnarray}
{\mathbf K}(\kappa)=\frac{1}{2}\ln(1-\kappa^2)+ O(1)
\end{eqnarray}
Thus, the low-field expansion of the electric polarization reveals the logarithmic non-analytical behavior:
\begin{eqnarray}
p_{x}{\approx}\frac{1}{\pi}\left\{E_x\sin^2\theta {-}\frac{1}{4} E_x E_y^2\sin^2(2\theta)\ln| E_x E_y\sin(2\theta)|\right\},
\nonumber\\
p_{y}{\approx}\frac{1}{\pi}\left\{E_y\cos^2\theta {-}\frac{1}{4} E_x^2 E_y\sin^2(2\theta)\ln| E_x E_y\sin(2\theta)|\right\}.
\nonumber\\
\label{pol_ln}
\end{eqnarray}
In particular, as it is seen from these equations, there is a singularity in the fourth derivative of the ground-state energy $\partial^4 e_0/(\partial E_x^2 \partial E_y^2)$ (see Eq.(\ref{pol_ln})).

\begin{figure}[tbp]
 \begin{center}
  \includegraphics[width=0.8\columnwidth]{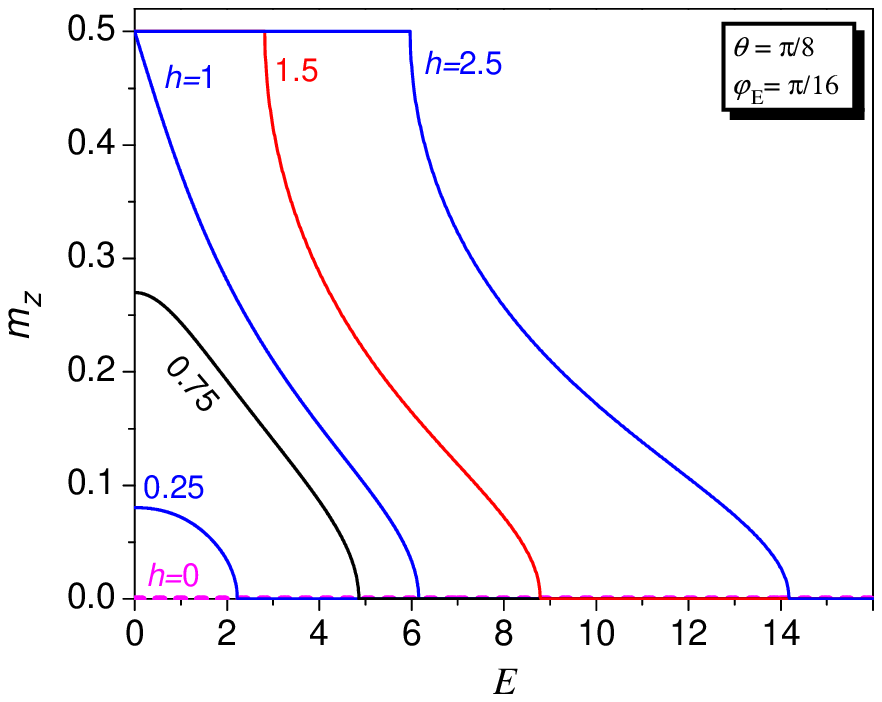}
  \includegraphics[width=0.8\columnwidth]{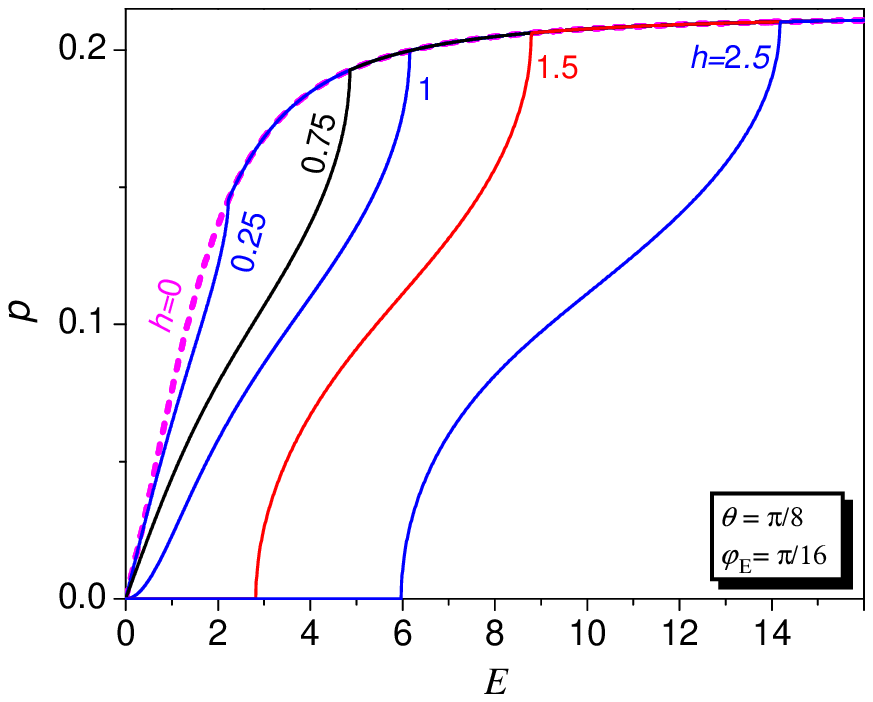}
  \includegraphics[width=0.8\columnwidth]{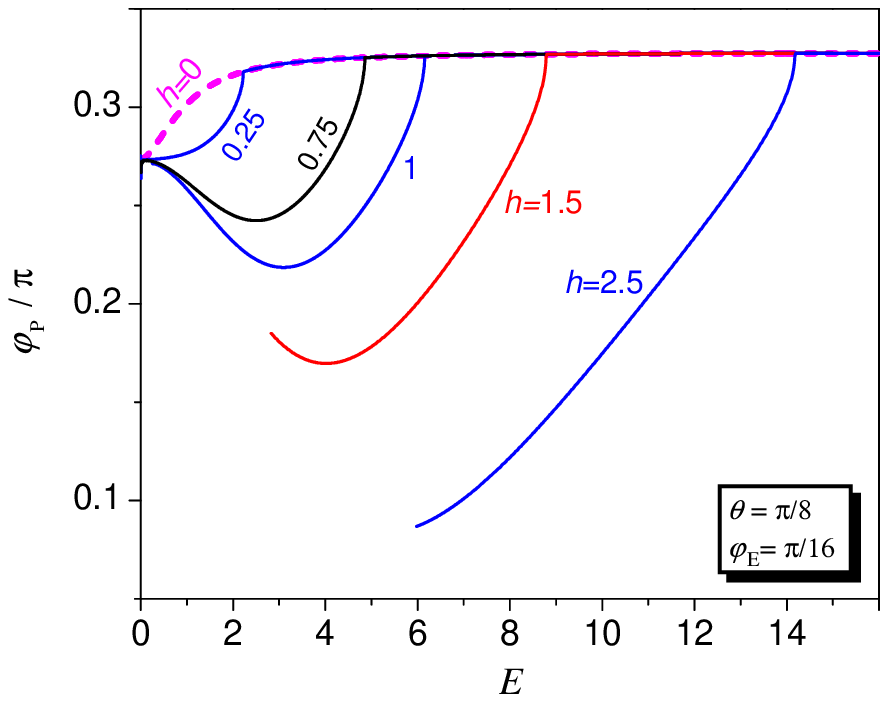}
 \end{center}
\caption{The magnetization (upper panel), the absolute value (middle panel) and the angle (lower panel) of the electric polarization vs  electric field for $\theta=\pi/8$, $\varphi^{}_{\rm E}=\pi/16$ and different magnetic fields $h=0,0.25,0.75,1,1.5,2.5$.}
\label{fig_el_pol}
\end{figure}

\begin{figure}[tbp]
 \begin{center}
  \includegraphics[width=0.8\columnwidth]{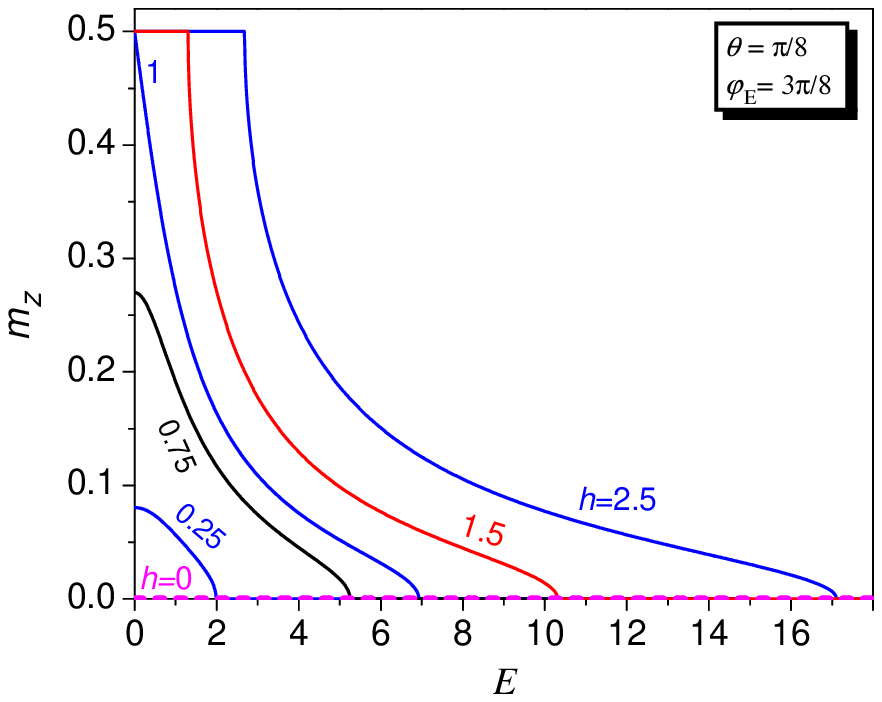}
  \includegraphics[width=0.8\columnwidth]{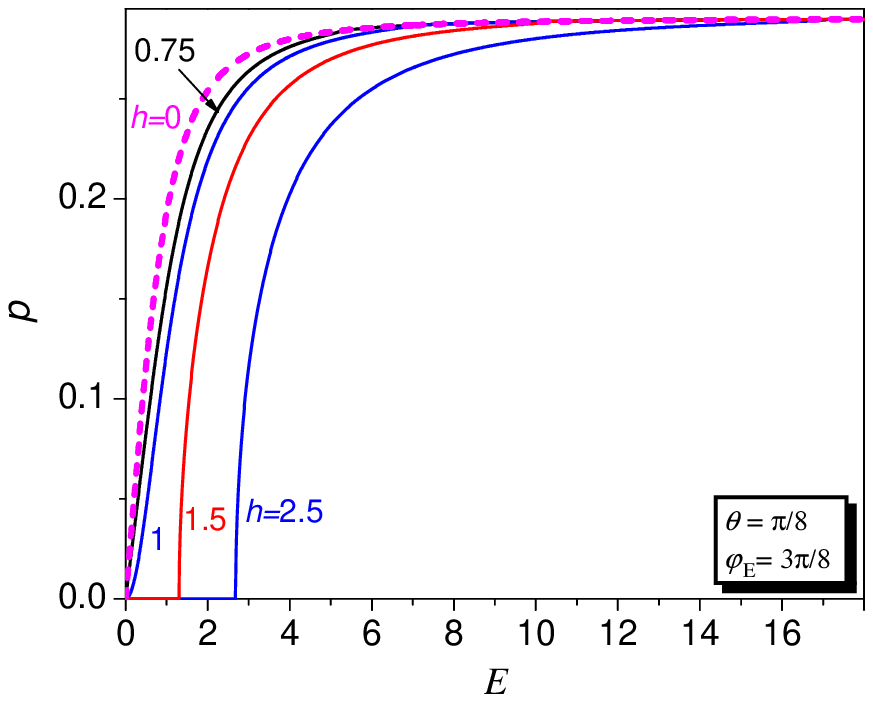}
  \includegraphics[width=0.8\columnwidth]{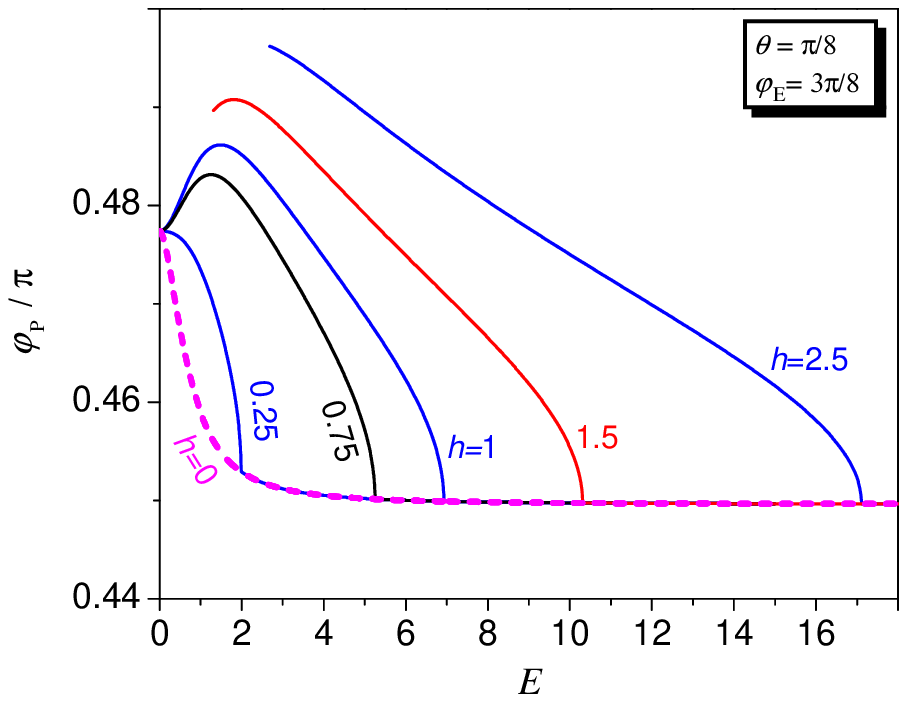}
 \end{center}
\caption{The magnetization (upper panel), the absolute value (middle panel) and the angle (lower panel) of the electric polarization vs electric field for $\theta=\pi/8$, $\varphi^{}_{\rm E}=3\pi/8$ and different magnetic fields $h=0,0.25,0.75,1,1.5,2.5$.}
\label{fig_el_pol2}
\end{figure}

 The electric-field dependence of the magnetization and polarization at $T=0$ is shown in Figs.~\ref{fig_el_pol},  \ref{fig_el_pol2}.
\begin{figure}[tbp]
 \begin{center}
  \includegraphics[width=0.8\columnwidth]{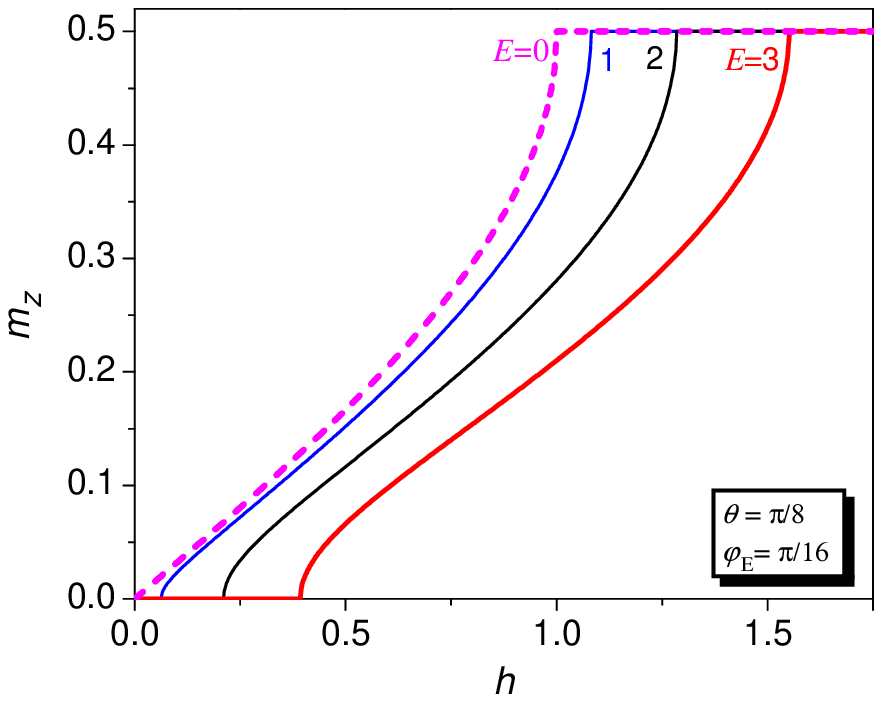}
  \includegraphics[width=0.8\columnwidth]{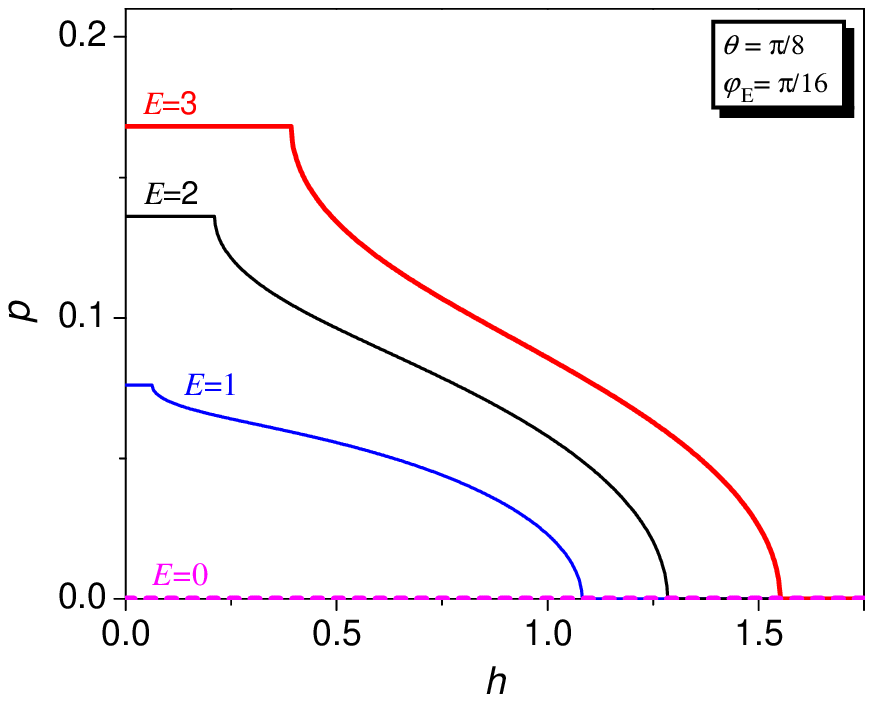}
  \includegraphics[width=0.8\columnwidth]{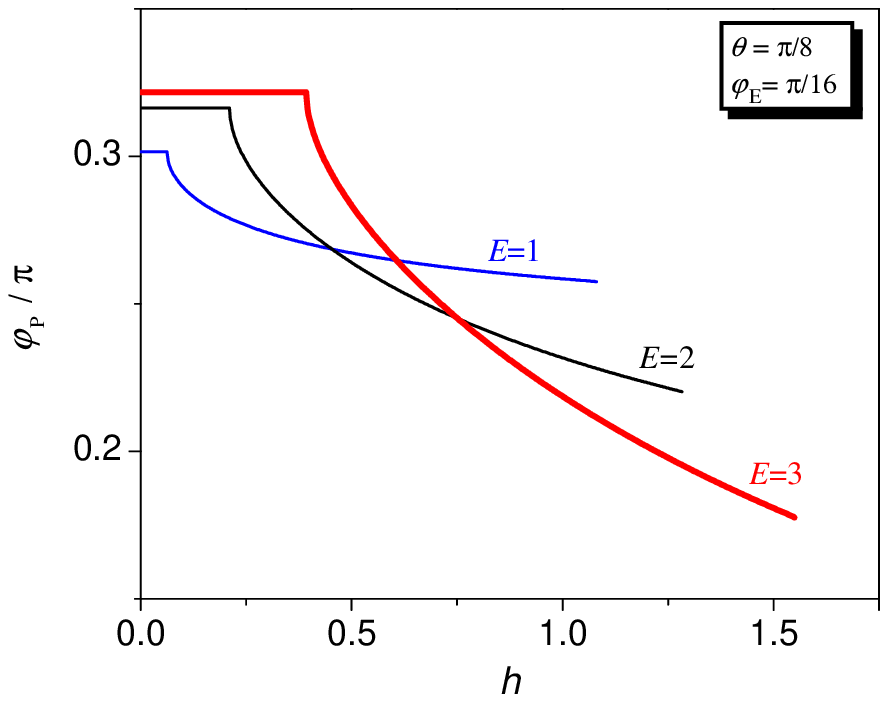}
 \end{center}
\caption{The magnetization (upper panel), the absolute value (middle panel) and the angle (lower panel) of the electric polarization vs magnetic field for $\theta=\pi/8$, $\varphi^{}_{\rm E}=\pi/16$ and different electric fields $E=0,1,2,3$.}
\label{fig_mag_pol}
\end{figure}
One can notice the demagnetizing effect of the electric field, when increasing field at first destroys the completely ordered phase and leads to the gapless spin-liquid phase. At the end, the gapped phase with zero magnetization emerges. Predictably, the electric polarization has opposite behavior. It starts from zero value in small electric fields passing through the spin-liquid phase to the  non-magnetic gapped phase. Interestingly, the electric polarization does not achieve the saturation value for a finite electric field. One can also noted, that when the magnetic field $h<J_+$ the polarization immediately emerges with the electric field with the linear law.

Fig.~\ref{fig_mag_pol} demonstrates the explicit dependence of the magnetization and polarization characteristics on the magnetic field. We see that the electric field applied in the $xy$ plain induces the gap in the excitation spectrum and the zero plateau in the magnetization curve. In the zero-plateau phase the polarization does not depend on the magnetic field according to Eq.~(\ref{gs_pol2}).

\subsection{The polarization angle}

It is useful to follow the behavior of the polarization angle $\varphi^{}_{\rm P}$
(between the $x$-axis and the vector of electric polarization)
in applied electric and magnetic fields (Figs.~\ref{fig_el_pol}--\ref{fig_mag_pol}).
It is worth mentioning, that if
$\varphi^{}_{\rm E} \in (0,\pi/2)$ and $\theta \in (0,\pi/4)$
and additionally $\varphi^{}_{\rm E}>\theta$
the polarization angle is larger than
$\theta$ at any magnitudes of electric and magnetic fields
while $\varphi^{}_{\rm P}$ can be larger or smaller than
$\varphi^{}_{\rm E}$ depend on values of parameters
$\theta$, $h$, $E$.
In general case the angle of ${\bf P}$
can be greater or less both than $\theta$ and
than $\varphi^{}_{\rm E}$.
Obviously, $\varphi^{}_{\rm P} \in [0,\pi/2]$ at $\varphi^{}_{\rm E} \in [0,\pi/2]$.
It is also interesting that
in the case $0<\varphi^{}_{\rm E}<\theta$ the polarization angle
$\varphi^{}_{\rm P}$ is increasing function of the strength of the electric field
at small enough and sufficiently large values of $h$,
while at intermediate values of magnetic field $\varphi^{}_{\rm P}(E)$
is non-monotonous function with one minima.
In the case $\theta <\varphi^{}_{\rm E}<\pi/2$ the behaviour of the polarization angle
is reverse: $\varphi^{}_{\rm P}(E)$ is decreasing or
non-monotonous function with one maxima.
The magnetic field dependence of $\varphi^{}_{\rm P}$ is affected by the relation between $\varphi^{}_{\rm E}$ and $\theta$.
In the spin-liquid phase $\varphi^{}_{\rm P}(h)$
is decreasing (increasing) function
at $0<\varphi^{}_{\rm E}<\theta$ ($\theta<\varphi^{}_{\rm E}<\pi/2$).

If the electric field
is directed along some of bonds (e.g., $\varphi^{}_{\rm E}=\theta$),
$\varphi^{}_{\rm P}$ does not depend on the strength of the electric and magnetic fields.
It can be shown explicitly calculating $\tan\varphi^{}_{\rm P}=p_y/p_x$.
Using Eqs. (\ref{gs_pol1}) and (\ref{gs_pol2}), we can easily get the expression:
\begin{eqnarray}\label{ang}
\tan\varphi^{}_{\rm P}=\frac{1+\xi(E,\theta)}{1-\xi(E,\theta)}\cot\theta,
\end{eqnarray}
where
\begin{eqnarray}\label{xi}
\xi(E,\theta)&=&
\frac{ E_-(1+E_+^2) \left({\mathbf E}(k_0|\kappa)-\frac{J_-}{J_+}{\mathbf F}(k_0|\kappa)\right)}
{ E_+(1+E_-^2) \left({\mathbf E}(k_0|\kappa)+\frac{J_-}{J_+}{\mathbf F}(k_0|\kappa)\right)} \nonumber
\end{eqnarray}
for the spin-liquid state ($|J_-|< h< J_+$), while one should put $k_0=\pi/2$ in case of the non-magnetic phase ($h<|J_-|$).
At $\varphi^{}_{\rm E}=\theta$ one has $E_-=0$ and
the Eq. (\ref{ang}) leads to the relation:
\begin{eqnarray}\label{tancot}
&& \tan\varphi^{}_{\rm P}=\cot\theta, \; {\rm if} \; \varphi^{}_{\rm E}=\theta.
\end{eqnarray}
As a consequence, the polarization is orthogonal to the direction of nonparallel bonds:
\begin{eqnarray}\label{varphi_E-theta}
&& \varphi^{}_{\rm P}= \frac{\pi}{2}-\theta, \; {\rm if} \; \varphi^{}_{\rm E}=\theta.
\end{eqnarray}
This simple result can be readily understood from the basic description of the model (\ref{301}). If the electric field and one type of bonds are collinear, the electric polarization cannot be created there due to the specific feature of KNB mechanism (\ref{201}). At the same time it induces a non-zero polarization perpendicular to the direction of other bonds. Such arguments are quite general and are valid for non-zero temperatures and more general $XXZ$ model as well, but cannot be applied to a model with the next-nearest neighbor interaction (see Appendix~\ref{app:equal_angles}).

There are several examples where the dependence of $\varphi^{}_{\rm P}$ on $\varphi^{}_{\rm E}$ are universal. In case of strong electric fields $E\to\infty$, the details of the exchange interactions become irrelevant. The system is governed exclusively by the electric field. Taking the limit $E\to\infty$ for fixed $\varphi^{}_{\rm E}$ in Eq.(\ref{gs_pol2}), we get:
\begin{eqnarray}
\label{E_infty}
\tan\varphi^{}_{\rm P}&=&-\frac{\tan\varphi^{}_{\rm E}}{\tan^2\theta}
\frac{{\mathbf E}(\kappa')-{\mathbf K}(\kappa')}
{{\mathbf E}(\kappa')-\left(\frac{\tan\varphi^{}_{\rm E}}{\tan\theta}\right)^2{\mathbf K}(\kappa')},
\nonumber\\
\kappa'&=&\lim_{E\to\infty}\kappa=\sqrt{1-\left(\frac{\tan\varphi^{}_{\rm E}}{\tan\theta}\right)^2},
\nonumber\\
&& {\rm if}\; \varphi^{}_{\rm E}<\theta;
\nonumber\\
\tan\varphi^{}_{\rm P}&=&-\frac{\tan\varphi^{}_{\rm E}}{\tan^2\theta}
\frac
{{\mathbf E}(\kappa'')-\left(\frac{\tan\theta}{\tan\varphi^{}_{\rm E}}\right)^2{\mathbf K}(\kappa'')}
{{\mathbf E}(\kappa'')-{\mathbf K}(\kappa'')},
\nonumber\\
\kappa''&=&\lim_{E\to\infty}\kappa=\sqrt{1-\left(\frac{\tan\theta}{\tan\varphi^{}_{\rm E}}\right)^2},
\nonumber\\
&& {\rm if}\; \varphi^{}_{\rm E}> \theta.
\end{eqnarray}
This result shows that such a dependence characterizes only the geometry of the lattice but not the spin model itself. It is useful also to estimate the dependence when the electric field is directed closely to the $x$- or $y$-axes.
Thus, at $E\to\infty$ and $\theta\ne0$ we have:
\begin{eqnarray}
\tan\varphi^{}_{\rm P}&=&-\frac{\tan\varphi^{}_{\rm E}}{\tan^2\theta}
\ln\left|\frac{\tan\varphi^{}_{\rm E}}{\tan\theta} \right|, \;{\rm if} \;\varphi^{}_{\rm E}\rightarrow 0;
\nonumber\\
\cot\varphi^{}_{\rm P}&\!=\!&
-\frac{\tan\varphi^{}_{\rm E}}{\cot^2\theta}
\ln\left|\frac{\tan\varphi^{}_{\rm E}}{\cot\theta}
 \right|,
\; {\rm if} \;\varphi^{}_{\rm E}\rightarrow \frac{\pi}{2}.
\label{E_infty2}
\end{eqnarray}

In case of low fields $E \to 0$, we can keep first terms in Eq.(\ref{pol_ln}) to get
\begin{eqnarray}
\label{E_low}
\tan\varphi^{}_{\rm P}&=&\frac{\tan\varphi^{}_{\rm E}}{\tan^2\theta}.
\end{eqnarray}
In Appendix~\ref{app:angles_small_fields} we showed that this result is also valid for $XXZ$ magnetoelectric on a zigzag chain.

\begin{figure}[tbp]
 \begin{center}
  \includegraphics[width=0.8\columnwidth]{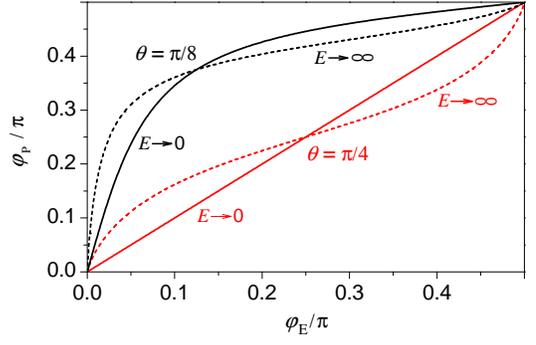}
 \end{center}
\caption{The polarization angle $\varphi^{}_{\rm P}$ as a function of the electric field angle $\varphi^{}_{\rm E}$ for infinitesimal small (solid line) and infinite (broken line) electric field at $\theta=\pi/8,\pi/4$.}
\label{fig_phi_P}
\end{figure}

The results for the polarization angle can be seen in Fig.~\ref{fig_phi_P}.
Both Eqs.~(\ref{E_infty}), (\ref{E_low}) recovers the limit $\varphi^{}_{\rm E}=\theta$, and, as one can see, the corresponding curves cross at this point.
For the small enough values of $\theta$
the polarization angle is larger than $\varphi^{}_{\rm E}$ both at $E\to0$
and at $E\to\infty$.
For $\theta\to \pi/4$ one has $\varphi^{}_{\rm P}\approx \varphi^{}_{\rm E}$
at $E\to0$, while at $E\to\infty$
the polarization angle is larger (smaller) than $\varphi^{}_{\rm E}$
if $\varphi^{}_{\rm E}<\pi/4$ ($\varphi^{}_{\rm E}>\pi/4$).

\subsection{Susceptibilities}

There are three kind of susceptibilities in our system, electric, magnetic and mixed, magnetoelectric.

The expressions for the electric susceptibilities per site
$\chi_{\mu\nu}= \frac{\partial p_{\mu}}{\partial E_{\nu}}$ ($\mu=x,y$)
can be found by a straightforward calculation. In the saturated phase they equal zero.
In the spin-liquid state (where $k_0$ is given in Eq.~(\ref{f_point}))
and in the non-magnetic phase (where $k_0=\pi/2$) we get:
\begin{eqnarray}
\label{gs_xi}
\chi_{\mu\nu}&=&\frac{1}{\pi}\Big\{
\frac{\partial^2 J_+}{\partial E_{\mu}\partial E_{\nu}} {\mathbf E}(k_0|\kappa)
\\ && \nonumber
+\Big[ \frac{h}{J_+} \frac{\partial J_+}{\partial E_{\mu}} +
\frac{h^2-J_+^2}{2h\kappa^2} \frac{\partial \kappa^2}{\partial E_{\mu}} \Big]
\frac{\partial k_0}{\partial E_{\nu}}
\\ && \nonumber
+\frac{J_+}{4\kappa^2(1\!-\!\kappa^2)} \Big[ \frac{J_+}{h}\sin k_0 \cos k_0 \!-\! {\mathbf F}(k_0|\kappa)\Big]
\frac{\partial \kappa^2}{\partial E_{\mu}} \frac{\partial \kappa^2}{\partial E_{\nu}}
\\ && \nonumber
+\frac{1}{2\kappa^2}
\Big({\mathbf E}(k_0|\kappa)-{\mathbf F}(k_0|\kappa)\Big)
\Big[ \frac{\partial J_+}{\partial E_{\mu}} \frac{\partial \kappa^2}{\partial E_{\nu}}
\\ && \nonumber
+ \frac{\partial J_+}{\partial E_{\nu}} \frac{\partial \kappa^2}{\partial E_{\mu}}
+ J_+ \frac{\partial^2 \kappa^2}{\partial E_{\mu} \partial E_{\nu}}
\\ && \nonumber
- J_+ \frac{2-\kappa^2}{2\kappa^2(1\!-\!\kappa^2)}
\frac{\partial \kappa^2}{\partial E_{\mu}} \frac{\partial \kappa^2}{\partial E_{\nu}}
\Big]
\Big\}.
\end{eqnarray}
Here
$\frac{\partial J_\pm}{\partial E_{\mu}}$,
$\frac{\partial \kappa^2}{\partial E_{\mu}}$ are given in Eqs.~(\ref{dJdE}),
(\ref{gs_pol1}),
$\frac{\partial k_0}{\partial E_{\mu}} =0$ in the non-magnetic phase, while
\begin{eqnarray}
\label{dq0dE}
\frac{\partial k_0}{\partial E_{\mu}} &=&
\frac{J_+(h^2{-}J_-^2)\frac{\partial J_+}{\partial E_{\mu}}
{+}J_-(J_+^2{-}h^2)\frac{\partial J_-}{\partial E_{\mu}}}
{(J_+^2{-}J_-^2)\sqrt{(J_+^2{-}h^2)(h^2{-}J_-^2)}}
\end{eqnarray}
in the spin-liquid state.
The derivatives used in the Eqs.~(\ref{gs_xi}) and(\ref{dq0dE}) were given by the following expressions:
\begin{eqnarray}
\label{diff expression}
\frac{\partial^2 \kappa^2}{\partial E_{\mu}\partial E_{\nu}} &=&
\frac{2J_-}{J_+^2} \Big[ \frac{J_-}{J_+}
\frac{\partial^2 J_+}{\partial E_{\mu}\partial E_{\nu}}
-\frac{\partial^2 J_-}{\partial E_{\mu}\partial E_{\nu}}
\Big]
\\  \nonumber &&
+
\frac{4J_-}{J_+^3} \Big[
\frac{\partial J_+}{\partial E_{\mu}} \frac{\partial J_-}{\partial E_{\nu}}
+
\frac{\partial J_-}{\partial E_{\mu}} \frac{\partial J_+}{\partial E_{\nu}}
\Big]
\\  \nonumber &&
-
\frac{2}{J_+^2} \Big[
\frac{\partial J_-}{\partial E_{\mu}} \frac{\partial J_-}{\partial E_{\nu}}
+
\frac{3 J_-^2}{J_+^2}
\frac{\partial J_+}{\partial E_{\mu}} \frac{\partial J_+}{\partial E_{\nu}}
\Big]
\\  \nonumber
\frac{\partial^2 J_{\pm}}{\partial E_{x}^2}&=&
{B_{\pm}}\sin^2\theta , \quad
\frac{\partial^2 J_{\pm}}{\partial E_{y}^2}=
{B_{\pm}}\cos^2\theta ,
\\  \nonumber
\frac{\partial^2 J_{\pm}}{\partial E_{x}\partial E_{y}}&=&
{B_{\pm}}\sin\theta\cos\theta ,
\\  \nonumber
B_{\pm}&=& \frac{1}{2\left(1+E_+^2\right)^{3/2}}
\pm \frac{1}{2\left(1+E_-^2\right)^{3/2}}
.
\end{eqnarray}

\begin{figure}[h]
 \begin{center}
  \includegraphics[width=0.8\columnwidth]{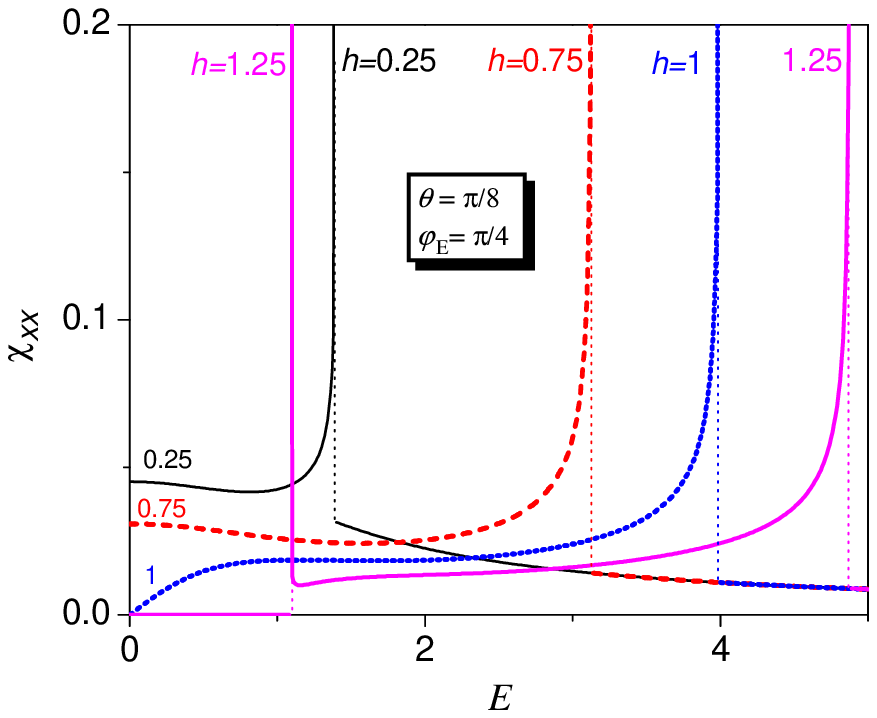}
  \includegraphics[width=0.8\columnwidth]{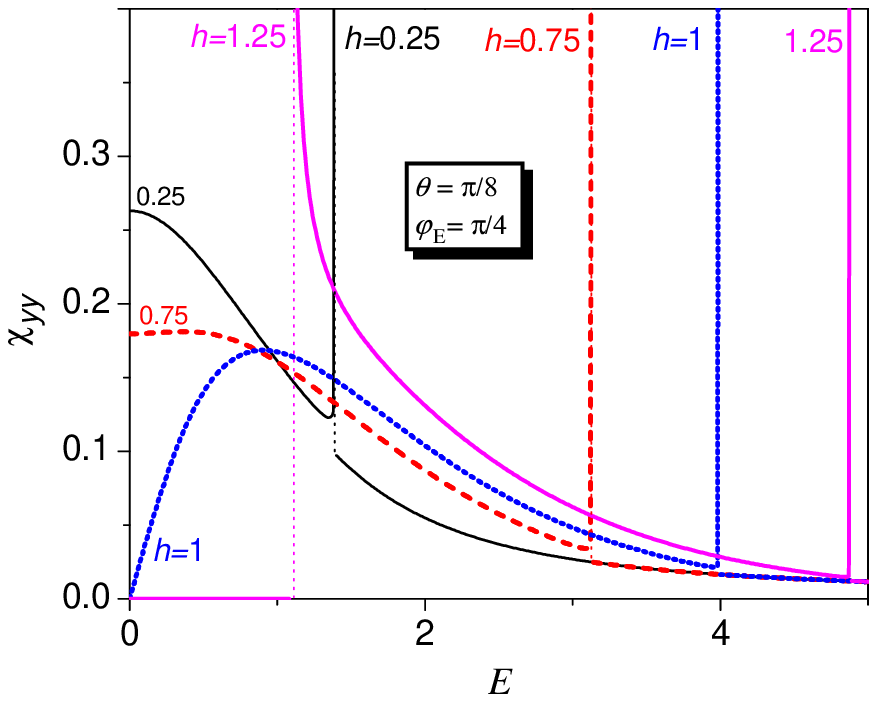}
  \includegraphics[width=0.8\columnwidth]{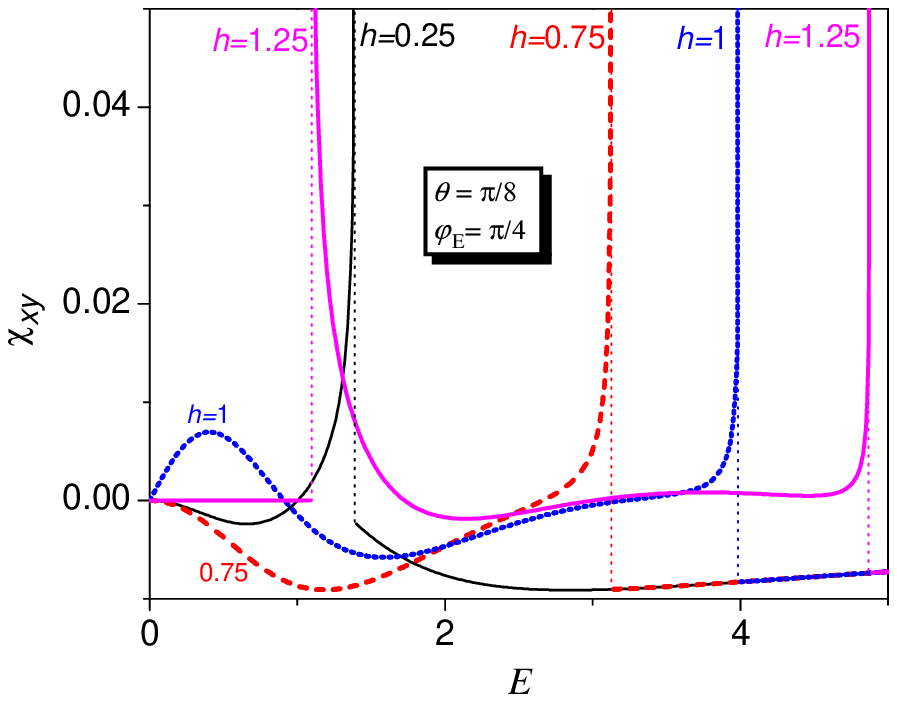}
 \end{center}
\caption{The electric susceptibilities $\chi_{xx}$, $\chi_{yy}$, and $\chi_{xy}$
at $T=0$ as a function of the electric field
for $\theta=\pi/8$, $\varphi^{}_{\rm E}=\pi/4$
and different values of $h$.}
\label{fig_chi-e}
\end{figure}

\begin{figure}[h]
 \begin{center}
  \includegraphics[width=0.8\columnwidth]{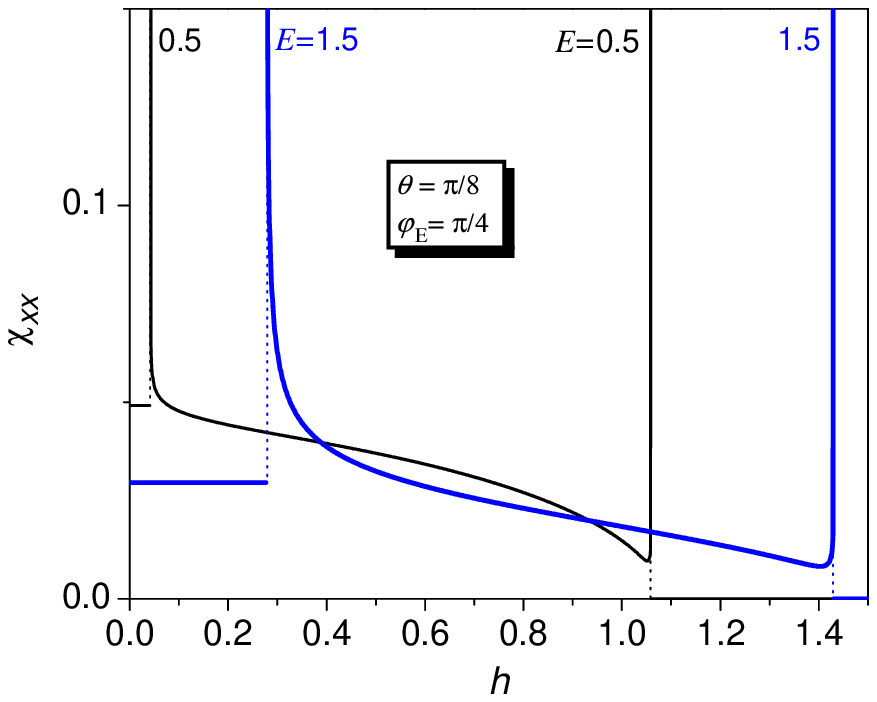}
  \includegraphics[width=0.8\columnwidth]{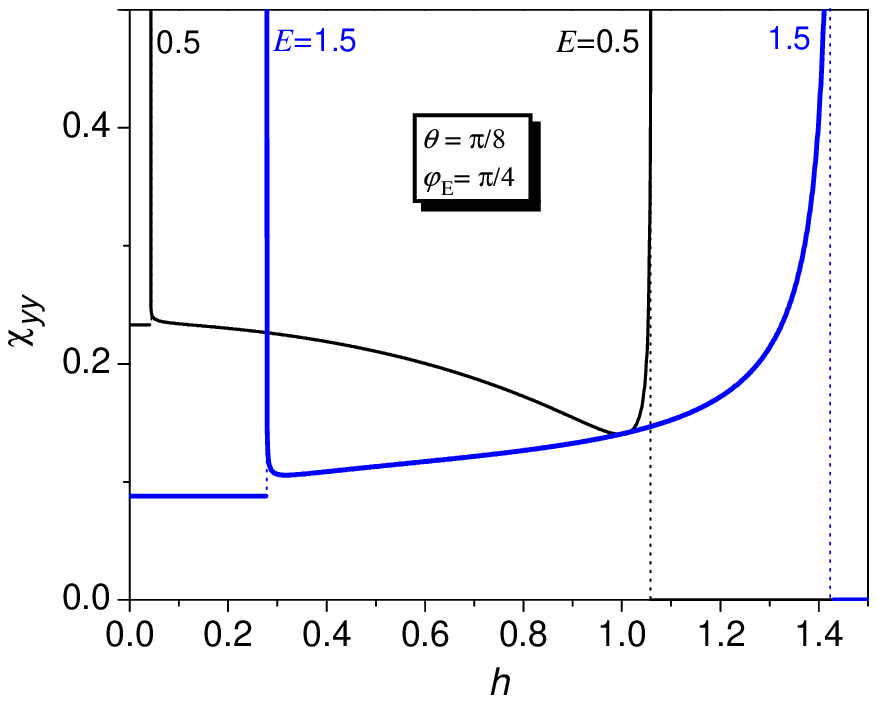}
  \includegraphics[width=0.8\columnwidth]{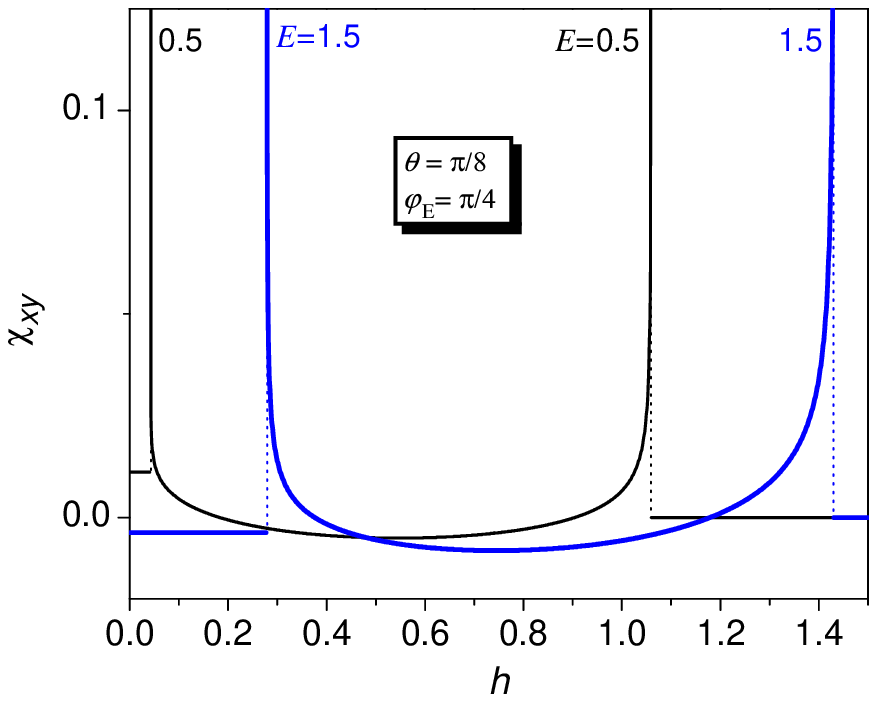}
 \end{center}
\caption{The electric susceptibilities $\chi_{xx}$, $\chi_{yy}$, and $\chi_{xy}$
at $T=0$ as a function of the magnetic field
for $\theta=\pi/8$, $\varphi^{}_{\rm E}=\pi/4$
and different values of $E$.}
\label{fig_chi-h}
\end{figure}

It is obvious that electric susceptibilities exhibit van Hove singularities
along the boundaries of the spin-liquid phase.

The most representative plots for the case when $\theta$ is substantially smaller than
$\varphi^{}_{\rm E}$ (e.g. $\theta=\pi/8$, $\varphi^{}_{\rm E}=\pi/4$)
for the electric (or magnetic) field dependence of the components of the
zero-temperature electric susceptibility of the system, diagonal
($\chi_{xx}$ and $\chi_{yy}$) and off-diagonal ($\chi_{xy}$),
are presented in the Fig. \ref{fig_chi-e} (or Fig. \ref{fig_chi-h}).
The singular peaks at the points of the quantum phase transitions are well pronounced here.

\begin{figure}[h]
 \begin{center}
  \includegraphics[width=0.8\columnwidth]{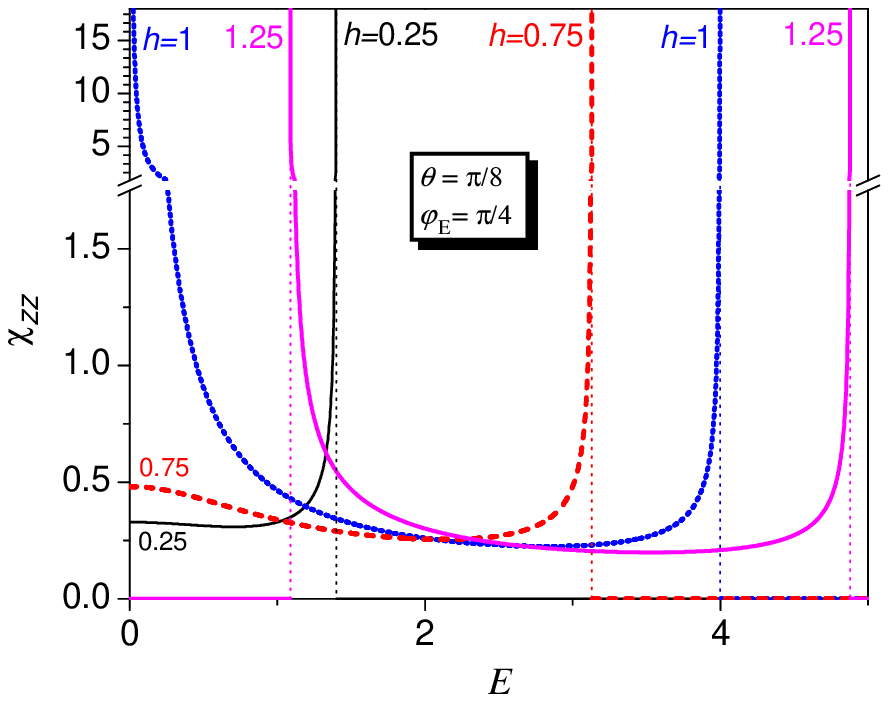}
  \includegraphics[width=0.8\columnwidth]{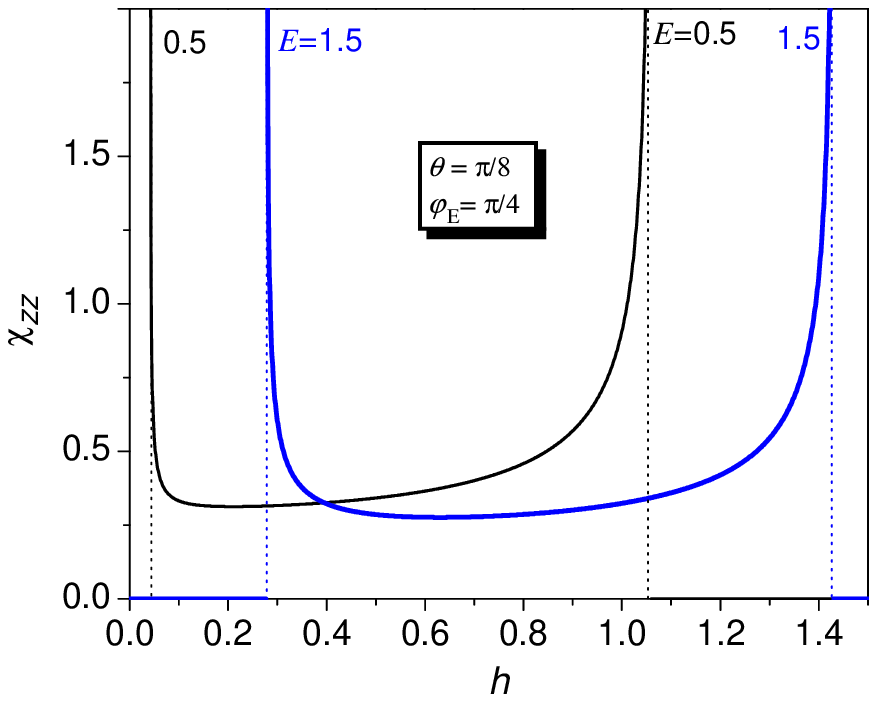}
 \end{center}
\caption{The magnetic susceptibility $\chi_{zz}$
at $T=0$ as a function of the electric field
(upper panel) and the magnetic field (lower panel)
for $\theta=\pi/8$, $\varphi^{}_{\rm E}=\pi/4$.}
\label{fig_chi-zz}
\end{figure}

The magnetic susceptibility per site
$\chi_{zz}= \frac{\partial m_z}{\partial h}$
in the spin-liquid state is quite simple:
\begin{eqnarray}
\label{gs_xi-zz}
\chi_{zz}&=&\frac{1}{\pi}
\frac{h}{\sqrt{(J_+^2-h^2)(h^2-J_-^2)}},
\end{eqnarray}
while in the saturated and in the non-magnetic phases it equals zero.
The corresponding plots of the $T=0$ magnetic susceptibility exhibiting the peaks pointing to the critical values of the electric and magnetic field can be found in Fig. \ref{fig_chi-zz} for $\theta=\pi/8$ and $\varphi^{}_{\rm E}=\pi/4$.

%
%

The magnetoelectric tensor components in the ground state can be obtained from the zero-temperature limit of Eq.~(\ref{me_tensor}), or by the direct taking of derivatives of Eq.~(\ref{gs_mag}),
\begin{eqnarray}
\label{me_tensor_gs}
\alpha_{z\mu}=
\left\{
\begin{array}{lll}
0, & {\rm if}\; h<|J_-|,\\
 - \frac{1}{\pi}\frac{\partial k_0}{\partial E_{\mu}}, & {\rm if}\: |J_-|{\le}h{\le}J_+,\\
0, & {\rm if}\; h>J_+,
\end{array}
\right.
\end{eqnarray}
where $\mu =x,y$, and the explicit form of the corresponding derivative of the Fermi momenta is given in the Eq. (\ref{dq0dE}).
It should be noted that the magnetoelectric tensor components as well as the
magnetic susceptibility equal zero outside the spin-liquid phase, while
electric susceptibilities equal zero in saturated phase only.

\begin{figure}[tbp]
 \begin{center}
  \includegraphics[width=0.8\columnwidth]{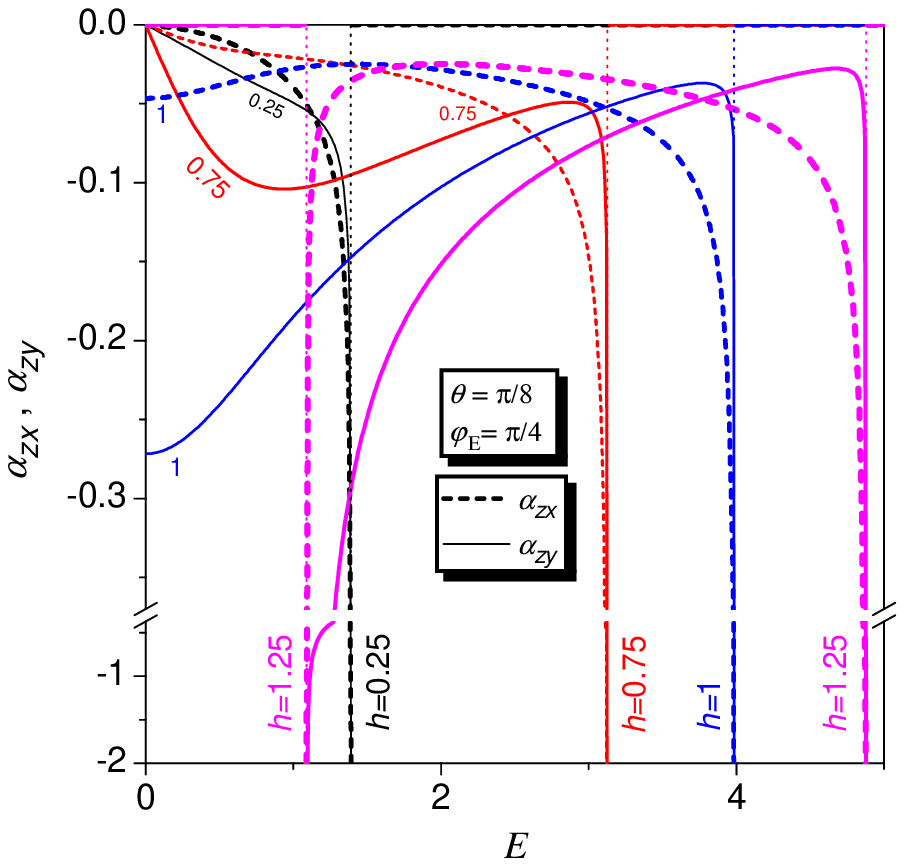}
  \includegraphics[width=0.8\columnwidth]{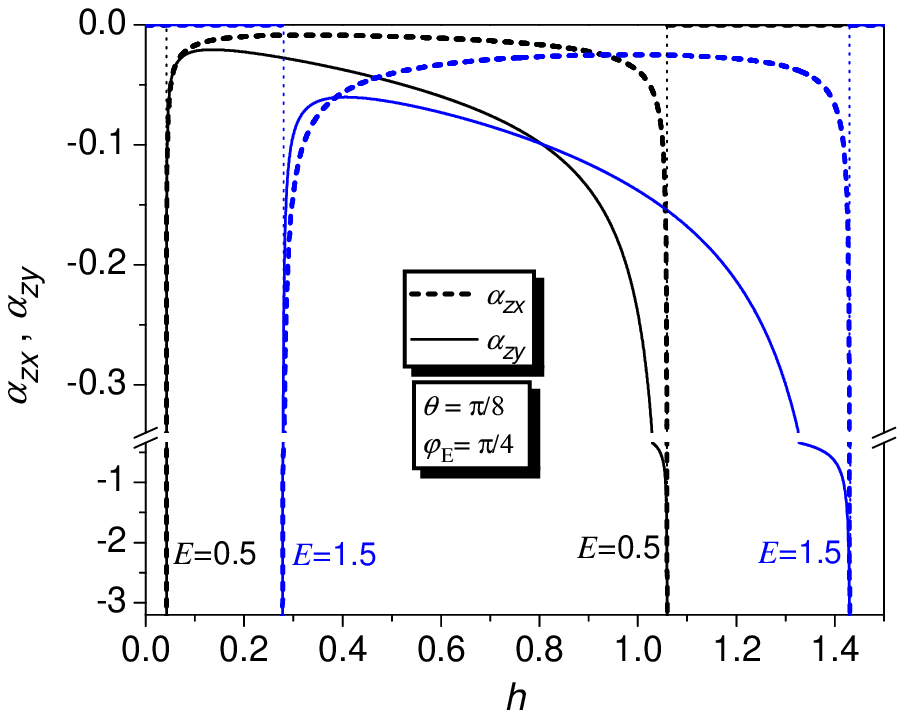}
 \end{center}
\caption{The magnetoelectric tensor at $T=0$ as a function of the electric field (upper panel) and the magnetic field (lower panel)
for $\theta=\pi/8$, $\varphi^{}_{\rm E}=\pi/4$.}
\label{fig_me_tensor}
\end{figure}

The corresponding plots of the zero temperature magnetoelectric tensor's non-zero
components dependence on electric and magnetic field are presented  in
Fig.~\ref{fig_me_tensor} for $\theta=\pi/8$ and $\varphi^{}_{\rm E}=\pi/4$. It is seen
that the magnetoelectric tensor exhibits a square-root van Hove singularities along
the boundaries of the spin-liquid phase.
The same feature can be occurred in the behavior of the magnetic and electric
susceptibilities (see Figs.~\ref{fig_chi-e}--~\ref{fig_chi-zz}).
It is interesting to observe that the magnetoelectric tensor is always zero for
$E=0$ except the case when the magnetic field takes its critical value.
This result follows directly from Eq.~(\ref{dq0dE}) in the limit $E\to 0$
at $h=J=1$.
It should be also noted that
in the case $h<1$ the
$zy$-component of the magnetoelectric tensor
at sufficiently small values of $h$ is decreasing
function of $E$  whereas
at  sufficiently large values of magnetic field it is non-monotonic function
(see Fig.~\ref{fig_me_tensor} at $h=0.25$, and $h=0.75$).

In the case when $\theta$ substantially larger than
$\varphi^{}_{\rm E}$ the results for ground state susceptibilities are somewhat different.
For example (see Figs. \ref{fig_chi-e},  \ref{fig_me_tensor}),
at $\theta<\varphi^{}_{\rm E}$ the curves of $\chi_{xx}(E)$ and $\alpha_{zx}(E)$
demonstrate
more sharpen behavior near the saturated phase
than near the non-magnetic one in the spin-liquid phase,
while the curves of
$\chi_{yy}(E)$ and $\alpha_{zy}(E)$
have more sharpen course
near the non-magnetic phase than near the saturated one.
At $\theta>\varphi^{}_{\rm E}$ we have the opposite situation.

\section{Thermodynamics}
\begin{figure}[tbp]
 \begin{center}
  \includegraphics[width=0.8\columnwidth]{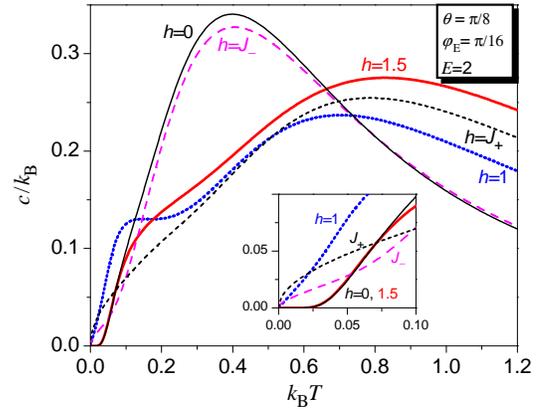}
 \end{center}
\caption{Specific heat as a function of temperature for $\theta=\pi/8$, $\varphi^{}_{\rm E}=\pi/16$, $E=2$ $h=0, J_{-},1, J_{+}, 1.5$.}
\label{fig_heat_t}
\end{figure}

In this Section we discuss the features of the temperature effect in the zigzag magnetoelectric. Let us start with the temperature-dependent specific heat which can show different behavior in various phases (see Fig.~\ref{fig_heat_t}).

In the gapped zero-magnetization and saturated phases we get the exponential asymptotic in the low-temperature specific heat, while the spin-liquid phase is characterized by a power-law dependence on temperature.\cite{knolle18}
In our case we get the linear dependence on the temperature for $|J_{-}|<h<J_{+}$ that can be clearly seen on the inset of Fig.\ref{fig_heat_t}. The most interesting case is the boundary of the spin-liquid phase ($h=J_{\pm}$) where the fermionic excitation spectrum touches zero, and, therefore, low-energy excitations gain quadratic dispersion. It results in the square root dependence of the specific heat $c\sim\sqrt{T}$.

\begin{figure}[!b]
 \begin{center}
  \includegraphics[width=0.8\columnwidth]{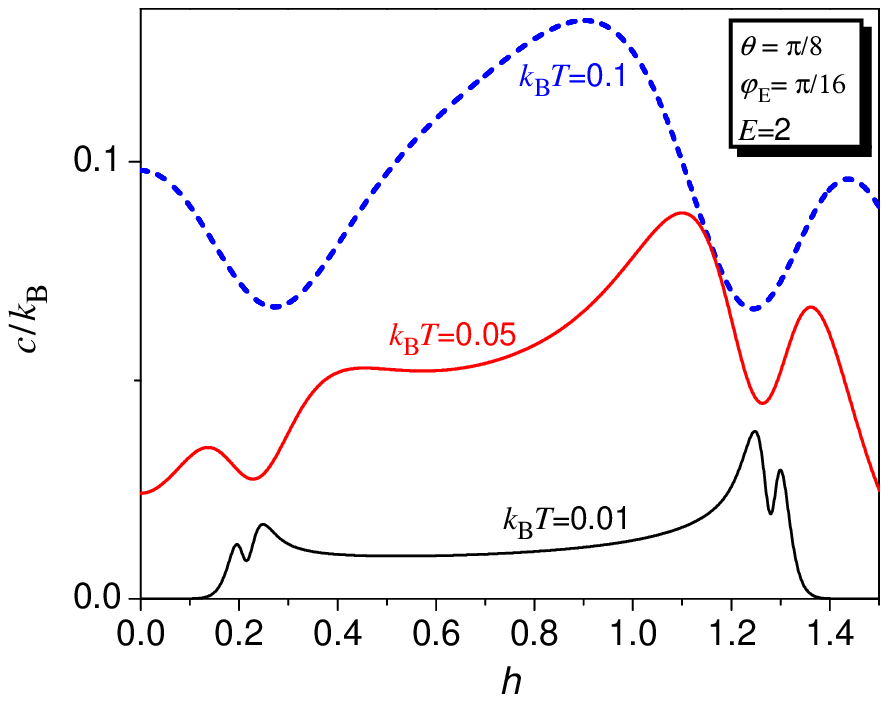}
  \includegraphics[width=0.8\columnwidth]{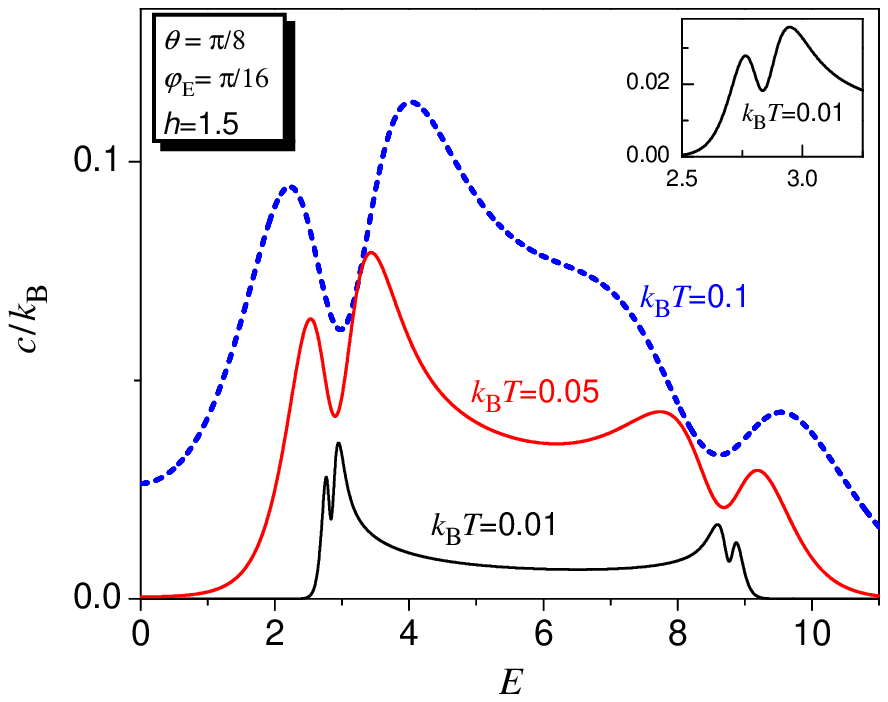}
 \end{center}
\caption{Specific heat as a function of the magnetic field (upper panel) for $E=2$, and of the electric field (lower panel) for $h=1.5$ and
$\theta=\pi/8$, $\varphi^{}_{\rm E}=\pi/16$, $k_{\rm B}T=0.01,0.05, 0.1$.}
\label{fig_heat}
\end{figure}

The field-dependent specific heat is presented in Fig.~\ref{fig_heat}.
We see that low-temperature curves signals the quantum phase transitions by a deep minima surrounding with two maxima in their vicinity.

\begin{figure}[tbp]
 \begin{center}
  \includegraphics[width=0.8\columnwidth]{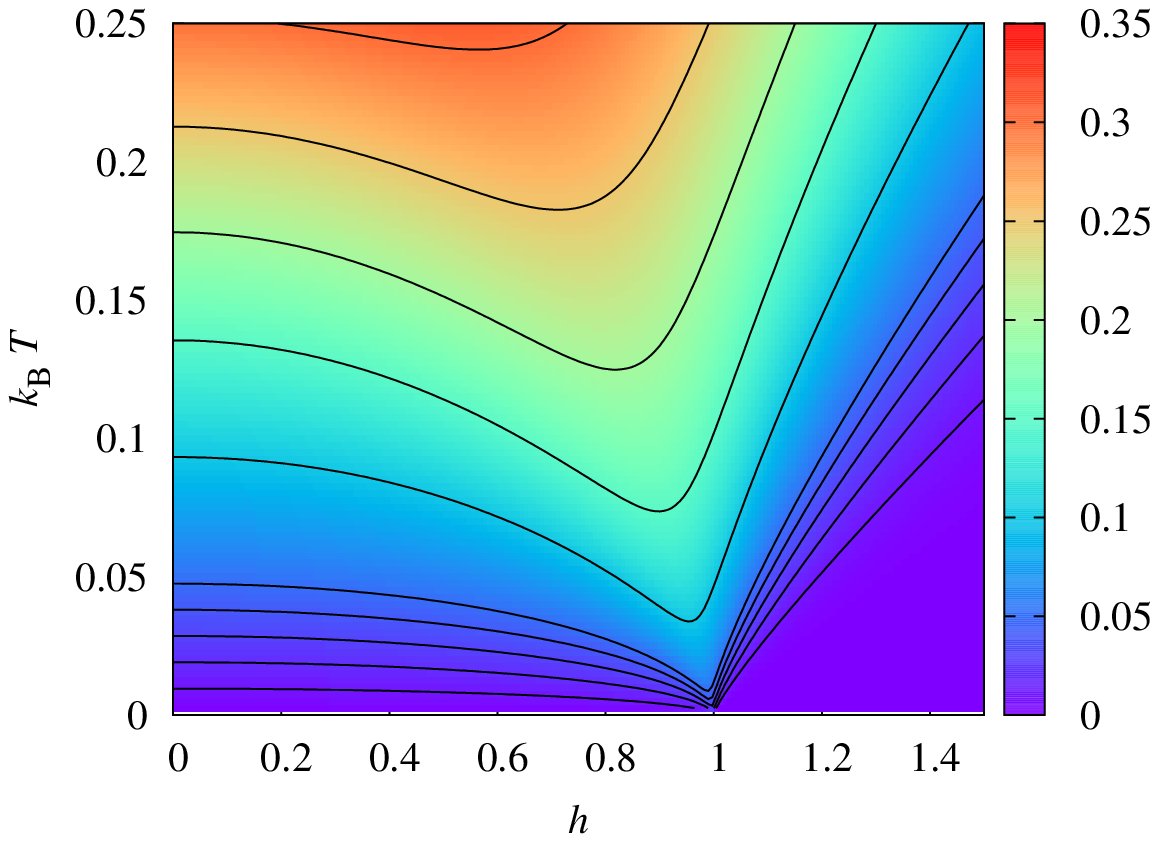}
  \includegraphics[width=0.8\columnwidth]{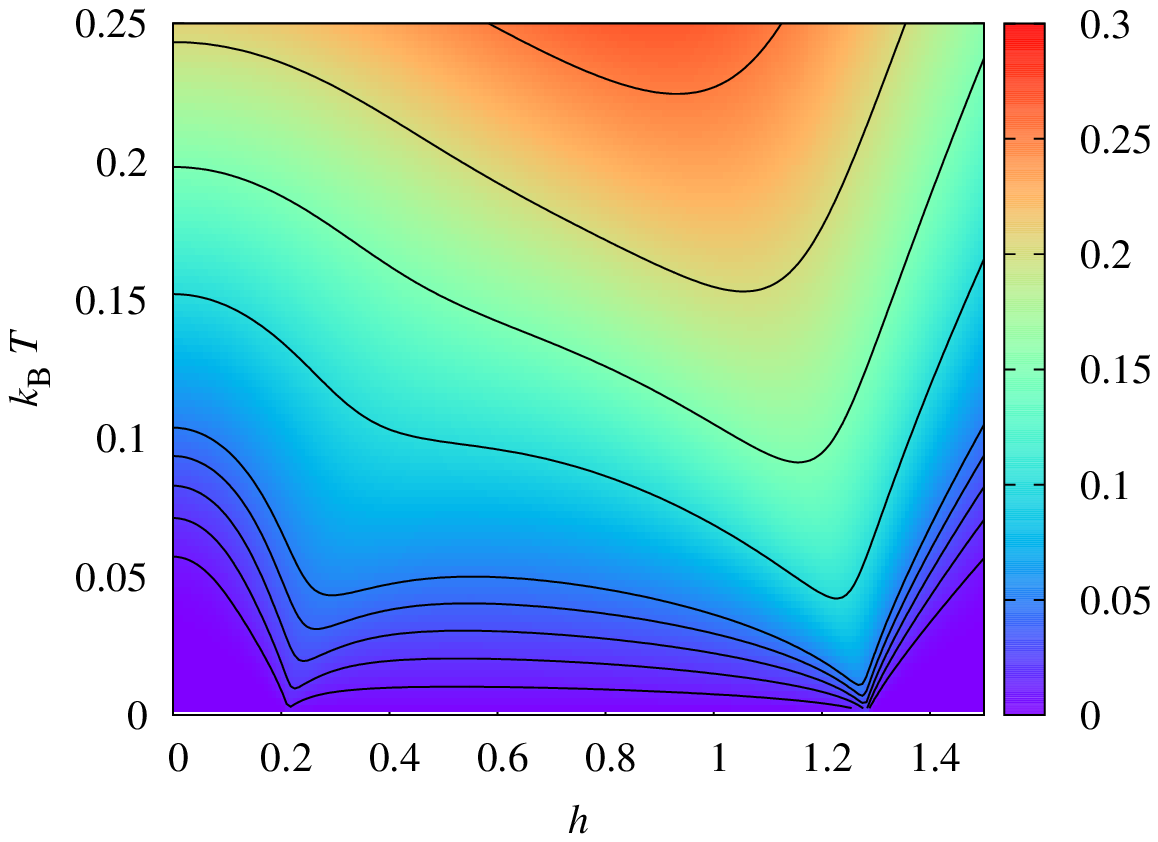}
 \end{center}
\caption{A density plot of the entropy as a function of the magnetic field and temperature for the zigzag chain with $\theta=\pi/8$, $\varphi^{}_{\rm E}=\pi/16$, $E=0$ (upper panel) and $E=2$ (lower panel). The curves with constant entropy correspond to
$S/k_{\rm B}= 0.01,0.02,0.03,0.04,0.05,0.1,0.15,0.2,0.25$.}
\label{fig_entr_den_h_T}
\end{figure}

\begin{figure}[tbp]
 \begin{center}
  \includegraphics[width=0.8\columnwidth]{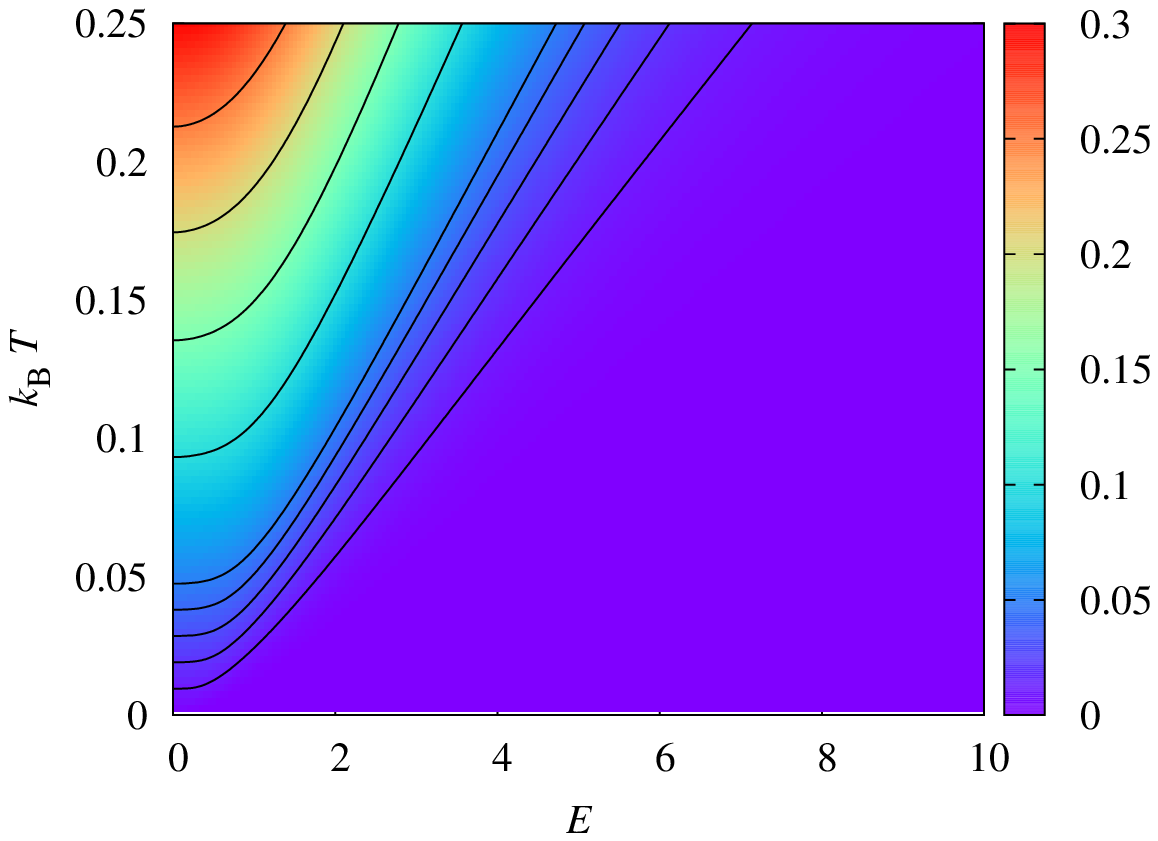}
  \includegraphics[width=0.8\columnwidth]{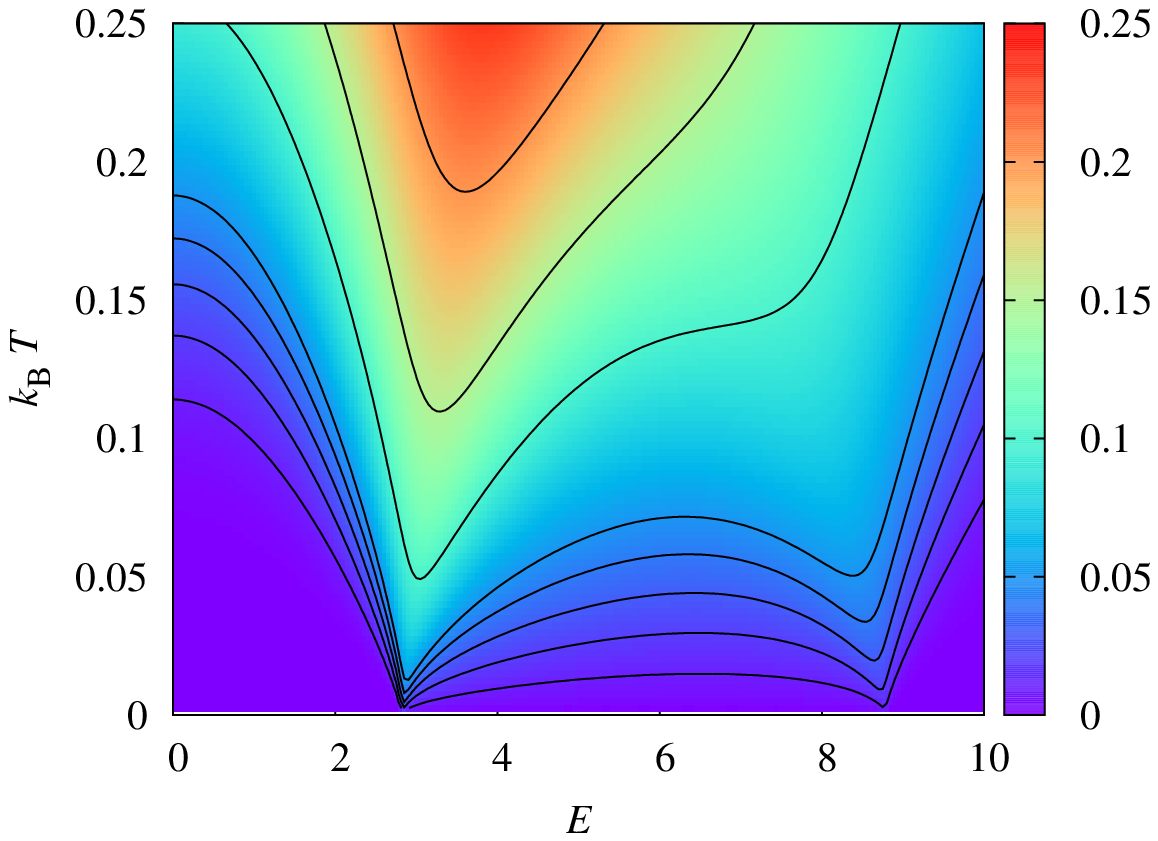}
 \end{center}
\caption{A density plot of the entropy as a function of the electric field $E$ and temperature for the zigzag chain with $\theta=\pi/8$, $\varphi^{}_{\rm E}=\pi/16$, $h=0$ (upper panel) and $h=1.5$ (lower panel). The curves with constant entropy correspond to
$S/k_{\rm B}= 0.01,0.02,0.03,0.04,0.05,0.1,0.15,0.2,0.25$.}
\label{fig_entr_den_E_T}
\end{figure}

The quantum spin paramagnets can be also attractive with respect to the enhanced magnetocaloric effect near critical fields.\cite{zhito07,tri10,top12,wolf14,wolf11,lang13}
Here, the inclusion of the electric field provides an additional possibility to tune this effect as well as opens an opportunity to study an electrocaloric effect.
The density plot for the entropy as a function of the magnetic field is shown in Fig.~\ref{fig_entr_den_h_T}. For the case of the vanishing electric field, we get a simple $XY$ chain which is known to show an enhanced magnetocaloric effect near the saturation field (see e.g. Ref [\onlinecite{zhito07}]). The application of the electric field in between $x$- and $y$-axis opens the gap between two fermionic bands and leads to the additional field-driven quantum phase transition at low field. It is reflected by a steep slope of the isentropes at comparably small field. Therefore, the application of the electric field creates a possibility to govern the strength of the magnetocaloric effect in this case.
The effect of the electric field on the entropy is also interesting to follow (see Fig.~\ref{fig_entr_den_E_T}).
We see that even for a small electric field the lines of constant entropy shows a strong dependence on it. The application of the magnetic field changes the character of the dependence from heating to freezing.

\section{Conclusions}
\label{sec6}

In the present paper we present the rigorous consideration of the effect of the
zigzag geometry on the properties of the quantum spin-$\frac{1}{2}$ $XY$
magnetoelectric chain within the KNB mechanism. By virtue of the interplay of the
geometry of the exchange interaction of the spins and the physics of the KNB mechanism
the dielectric behavior of the system exhibits anisotropy. Besides the usual off-diagonal
components of the magnetoelectric tensor, $\alpha_{zx}$ and $\alpha_{zy}$ the chain
possesses the diagonal $\chi_{xx}, \chi_{yy}$ and off-diagonal $\chi_{xy}$ components
of the electric susceptibility.

The ground-state phase diagram has been studied.
We revealed that the application of the electric field opens the gap in the excitation
spectrum and leads to the appearance of the gapped zero-plateau phase. In general,
the effect of the electric field is twofold: (i) it destroys the magnetic order, (ii) it
induces the gap in the spin-liquid phase.

We have also analyzed the direction of the
polarization angle caused by the electric field, and found that the applied magnetic
field decrease (increase) it in case the angle of the electric field $\varphi^{}_{\rm E}$
is smaller (larger) than $\theta$.
The behavior of the angle between polarization vector and the direction of the
chain (the $x-$axis) can be non-monotonous with respect to the magnitude
and the direction of the applied electric field. When the direction of the external
electric field is collinear with the chain bonds ($\varphi^{}_{\rm E}=\theta$) the direction
of the polarization is unaffected by the magnitudes of the electric and magnetic fields. Several cases of the universal dependence between the polarization and applied electric field angles are figured out in the limit of strong electric field.

In addition to the zero-temperature properties, we analyze some thermal effects, particularly the temperature behavior of the specific heat for various regimes, as well as its low-temperature electric and magnetic field dependence. The isentropes both in $(E, T)$ and $(h, T)$ planes are presented. The features of the magnetocaloric and electrocaloric effects are discussed.
We found out that the appearance of the quantum phase transition at a low magnetic field by the  application of the electric field is favorable for the enhanced magnetoelectric effect.

\section*{Acknowledgments}
The authors express their gratitude to Oleg Derzhko and Taras Krokhmalskii for the valuable discussions.
The present research was partially supported by the ICTP (OEA, network NT-04).
O.~B. and T.~V. acknowledge the kind hospitality of the Yerevan State University in 2016.
V.~O. also acknowledges the partial financial support form the grants
of the State Committee of Science of Armenia No.~13-1F343 and SFU-02. He also would like to thank the ICMP for warm hospitality during his visits to L'viv in 2016 and 2017.

\appendix

\section{Polarization angle for $\varphi^{}_{\rm E}=\theta$}
\label{app:equal_angles}
Let us consider a more general case of the $XXZ$ magnetoelectric on a zigzag chain defined by the Hamiltonian:
\begin{eqnarray}
\label{xxz_me}
&& {\mathcal H}={\mathcal H}_{xxz} + {\mathcal H}_{E},
\nonumber\\
&& {\mathcal H}_{xxz}=\sum_{j=1}^N(
J_{xy}(s^x_j s^x_{j+1} + s^y_j s^y_{j+1}) + J_z s^z_j s^z_{j+1}
-h{s}^z_j),
\nonumber\\
&& {\mathcal H}_{E}=\sum_{j=1}^N\left(E_y\cos\theta+(-1)^jE_x\sin\theta \right)D_{j,j+1}.
\end{eqnarray}
For the sake of simplicity
we introduced here the notation
$D_{j,j+1}=\left(s_j^x s_{j+1}^y-s_j^y s_{j+1}^x\right)$.

In the case $\varphi^{}_{\rm E}=\theta$, the electric field acts on the bonds which are non-collinear with the electric field:
\begin{eqnarray}
{\mathcal H}_{E}=\sum_{j=1}^{N/2}E\sin(2\varphi^{}_{\rm E})D_{2j,2j+1}.
\end{eqnarray}
Now, we use the standard notation for the thermal average:
\begin{eqnarray}
\langle D_{j,j+1}\rangle=\frac{{\rm Tr}\left\{D_{j,j+1}\exp\left[-\beta {\mathcal H}(E\sin(2\varphi^{}_{\rm E}))\right]\right\}}
{{\rm Tr}\exp\left[-\beta {\mathcal H}(E\sin(2\varphi^{}_{\rm E}))\right]}.
\end{eqnarray}
We can introduce the spatial inversion operator $I$, which sets the opposite ordering of sites.
Using the following relations $ID_{j,j+1}I=-D_{j,j+1}$, $I{\mathcal H}_{E}I=-{\mathcal H}_{E}$, $I{\mathcal H}_{xxz}I={\mathcal H}_{xxz}$, one can show that
\begin{eqnarray}
\langle D_{j,j+1}\rangle=-\frac{{\rm Tr}\left\{D_{j,j+1}\exp\left[-\beta {\mathcal H}(-E\sin(2\varphi^{}_{\rm E}))\right]\right\}}
{{\rm Tr}\exp\left[-\beta {\mathcal H}(-E\sin(2\varphi^{}_{\rm E}))\right]}
\end{eqnarray}
To invert the sign before the electric field, we apply the transformation (\ref{rotation_transf}) with the following parameters:
\begin{eqnarray}
\label{rot2}
\phi_{2j-1}=\phi_{2j}=2j\phi_0,
\nonumber\\
\tan\phi_0=E\sin(2\varphi^{}_{\rm E}),
\end{eqnarray}
thus, proving $\langle D_{2j-1,2j}\rangle=0$.
On the contrary $\langle D_{2j,2j+1}\rangle\neq 0$,
and the polarization components $p_x=\sin\theta\langle D_{2j,2j+1}\rangle$,
$p_y=-\cos\theta\langle D_{2j,2j+1}\rangle$
direct the polarization perpendicularly to the non-colinear bonds
(see Eq.~(\ref{varphi_E-theta})).
It should be noted that the given arguments are not valid for the quantum spin chain with the next-nearest-neighbor interaction, since the rotation (\ref{rot2}) effects the latter coupling.

\section{Polarization angle for small fields}
\label{app:angles_small_fields}
We consider again the Hamiltonian (\ref{xxz_me}) ${\mathcal H}(E_x\sin\theta,E_y\cos\theta)$.
To get the polarization angle, we can expand the polarizations in small fields:
\begin{eqnarray}
&&p_\mu=\chi_{\mu x}E_x + \chi_{\mu y}E_y,
\nonumber\\
&&\chi_{\mu\nu}{=}\frac{1}{N}\sum_{i,j}(\langle P_{i,i+1}^\mu P_{j,j+1}^\nu \rangle {-} \langle P_{i,i+1}^\mu\rangle \langle P_{j,j+1}^\nu \rangle),
\nonumber
\end{eqnarray}
thus,
\begin{eqnarray}
\label{varphi_P_xxz}
\tan\varphi^{}_{\rm P}=\frac{\chi_{y x}\cos\varphi^{}_{\rm E} + \chi_{y y}\sin\varphi^{}_{\rm E}}{\chi_{x x}\cos\varphi^{}_{\rm E} + \chi_{x y}\sin\varphi^{}_{\rm E}}
\end{eqnarray}
Let us introduce reduced quantities ${\tilde E}_x=E_x\sin\theta$, ${\tilde E}_y=E_y\cos\theta$,
$\tilde{P}_{i,i+1}^x=P_{i,i+1}^x/\sin\theta$, $\tilde{P}_{i,i+1}^y=P_{i,i+1}^y/\cos\theta$, and
\begin{eqnarray}
&&\tilde\chi_{\mu\nu}{=}\frac{1}{N}\sum_{i,j}(\langle \tilde{P}_{i,i+1}^\mu \tilde{P}_{j,j+1}^\nu \rangle
{-} \langle \tilde{P}_{i,i+1}^\mu\rangle \langle \tilde{P}_{j,j+1}^\nu \rangle).
\nonumber
\end{eqnarray}
It is easy to see that
$\chi_{xx}=\sin^2\theta\tilde\chi_{xx}$, $\chi_{yy}=\cos^2\theta\tilde\chi_{yy}$,
$\chi_{xy}=\sin\theta\cos\theta\tilde\chi_{xy}$, $\chi_{yx}=\sin\theta\cos\theta\tilde\chi_{yx}$.

Next, we are going to prove that $\tilde\chi_{xy}=\tilde\chi_{yx}=0$.
The Hamiltonian and $\sum_i\tilde{P}_{i,i+1}^y$ is invariant with respect to the translation, while $\sum_i\tilde{P}_{i,i+1}^x$ changes its sign after the one-site translation. It turns the correlation function $\sum_{i,j}\langle \tilde{P}_{i,i+1}^x \tilde{P}_{j,j+1}^y \rangle$ and the corresponding susceptibility $\tilde\chi_{xy}$ to zero.
Finally, it is easy to prove that $\tilde\chi_{xx}=\tilde\chi_{yy}$.
It follows from $f(E_x\sin\theta,E_y\cos\theta)=f(E_y\cos\theta,E_x\sin\theta)$, shown in sec. \ref{sec3}.
Inserting it into Eq.~(\ref{varphi_P_xxz}) we recover Eq.~(\ref{E_low}).

\end{document}